\documentclass[12pt,preprint]{imsart}
\RequirePackage[colorlinks,citecolor=blue,urlcolor=blue]{hyperref}
\usepackage{amssymb,amscd,amsthm, verbatim,amsmath,color,fancyhdr, mathrsfs}
\usepackage[authoryear]{natbib}
\usepackage{graphicx,tabularx,booktabs}
\usepackage{turnstile}
\usepackage{fancyhdr}
\usepackage{caption,subcaption}
\usepackage{url}
\usepackage{color,verbatim,enumerate}
\usepackage[normalem]{ulem}
\usepackage{marginnote} 
\usepackage{multirow}
\usepackage[letterpaper, left=2.5cm, right=2.5cm, top=2.5cm,
bottom=2.5cm,dvips]{geometry}
\usepackage{tikz}
\usepackage{bm}
\usepackage[ruled, 
lined, 
commentsnumbered]{algorithm2e}

\theoremstyle{plain} 
\newtheorem{theorem}{Theorem}[section]

\newtheorem{conj}{Conjecture}

\newtheorem{proposition}[theorem]{Proposition}
\newtheorem{prop}[theorem]{Proposition}
\theoremstyle{definition}
\newtheorem{defn}[theorem]{Definition}

\newtheorem{remark}[theorem]{Remark}
\newtheorem{example}[theorem]{Example}

\newcommand{\R}{\mathbb{R}}
\newcommand{\T}{\mathcal{T}}

\newcommand{\B}{\mathbb{B}}
\newcommand{\M}{\mathcal{M}}

\newcommand{\er}{Erd\"os-R\'enyi}
\newcommand{\blocks}{\mathcal Z}
\newcommand{\gof}{\mbox{GoF}_\blocks}
\newcommand{\gofz}{\mbox{GoF}_{\blocks_0}}

\newcommand{\gofzhatiterated}{\mbox{GoF}_{\hat\blocks_j}}
\renewcommand{\P}{\mathbb P}
\makeatletter
\newcommand\incircbin
{%
  \mathpalette\@incircbin
}
\newcommand\@incircbin[2]
{%
  \mathbin%
  {%
    \ooalign{\hidewidth$#1#2$\hidewidth\crcr$#1\bigcirc$}%
  }%
}

\makeatother

\DeclareMathOperator{\conv}{conv}

\newcommand*\samethanks[1][\value{footnote}]{\footnotemark[#1]}

\begin{document}
		\begin{frontmatter}
			\title{Monte Carlo goodness-of-fit tests for degree corrected and related stochastic blockmodels
			}
			\runtitle{Testing goodness of fit of  stochastic blockmodels}
			
			\begin{aug}
			\author[A]{ Vishesh Karwa\thanks{Lead author.}},
			\author[B]{ Debdeep Pati\samethanks},
			\author[c]{Sonja Petrovi\'c\samethanks},
			\author[d]{Liam Solus\samethanks}, 
			\author[e]{Nikita Alexeev},  
			\author[f]{Mateja Rai\v{c}},
			\author[g]{Dane Wilburne},
			\author[h]{ Robert Williams},
			\and
			\author[i]{Bowei Yan}
			\runauthor{Karwa et al.}

\address[A]{Temple University, Philadelphia, USA}
\address[B]{Texas A\& M University, College Station, USA}
\address[c]{Illinois Institute of Technology, Chicago, USA}
\address[d]{KTH Royal Institute of Technology in Stockholm, Sweden}
\address[e]{Independent Researcher, Tel Aviv, Israel}
\address[f]{Univ of Illinois at Chicago, USA}
\address[g]{Mitre corporation, USA}
\address[h]{Rose-Hulman institute of technology, USA}
\address[i]{Citadel, Chicago, USA}

\end{aug}

\begin{abstract}
We construct Bayesian and frequentist finite-sample goodness-of-fit tests for three different variants of the stochastic blockmodel for network data. 
Since all of the stochastic blockmodel variants are log-linear in form  when  block assignments are known, 
the tests for the \emph{latent} block model versions combine a block membership estimator with the algebraic statistics machinery for testing goodness-of-fit in log-linear models. We describe Markov bases and marginal polytopes of the variants of the stochastic blockmodel, and discuss how both facilitate the development of goodness-of-fit tests and understanding of model behavior. 

The general testing methodology developed here extends to any finite mixture of log-linear models on discrete data, and as such is the first application of the algebraic statistics machinery for latent-variable models.
\end{abstract}

\end{frontmatter}	
\small
\underline{Keywords:} algebraic statistics; goodness-of-fit tests; latent class models; Markov basis; networks; relational data; stochastic blockmodels. 
\normalsize



\section{Introduction}
\label{sec:introduction}

Exciting theoretical and algorithmic developments for analysis of networks have been motivated by the ever-increasing availability of relational data in diverse fields such as social sciences, web recommender systems, protein networks, genomics and neuroscience. 
There is a rich literature on probabilistic modeling of network data that stemmed from the classical \er\ random graphs introduced in  \cite{erdos1961evolution} and the contemporary article \cite{Gilbert}.  The exponential family approach to random graph modeling has its roots in \cite{holland1981exponential}, and has evolved through Markov graphs for modeling dependencies between dyads, \citep{frank1986markov}, leading to the general exponential family form (ERGMs), see \cite{Robins2007} for instance. A detailed review on the development and analysis of statistical network models is offered in \cite{goldenberg2010survey},  which points to a recent emphasis on stochastic blockmodels (SBMs). 
As a generalization of the \er\ model,  the SBM allows the probabilities of  occurrences of the edges between different pairs of nodes to be distinct, depending on the block membership of the two nodes in the pair.  Originally proposed in the social sciences  by \cite{fienberg1981categorical} for directed and undirected graphs, the SBM has since been extended to latent blocks in the undirected case in \cite{holland1983stochastic} (see also \cite{nowicki2001estimation}). Other extensions include latent space models \citep{hoff2002latent}, variable degree distribution models \citep{karrer2011stochastic},  SBMs for dynamically evolving networks \citep{fu2009dynamic,matias2015statistical},  and also SBMs that allow variable membership of the nodes \citep{airoldi2008mixed}. 
The SBM has taken center stage in computer science, statistics and machine learning as one of the more popular approaches that model and capture community structures in networks.

The question of model fit is an important and practical one,  not only relevant for the adequacy of a particular single model, but also for asking \emph{whether} we should fit a blockmodel   to the data in the first place. 
Traditionally, a large part of the literature on computation and modeling  does not  address model adequacy issues  beyond heuristic algorithms, detailed  to a general extent in \cite{HunterGoF} and \cite{carnegie2015approximation}.  This gap is largely due to   inherent model complexity or degeneracy and the lack of tools that can handle network models and sparse small-sample data. While there exists an understanding  of effective sample size in network data, as discussed in  \cite{Kolaczyk-EffectiveSampleSize}, which points to non-standard asymptotics in networks, most well-known goodness-of-fit tests are based on large sample approximations. Examples of inadequacy of the use of asymptotics  date as far back as \cite{Haberman88}, who illustrates why exact testing should be performed when sample size is small, or if some cell entries in a contingency table data are much smaller than others. That the sparse contingency table setting is directly relevant to network models is evident from \cite{Fienberg2012:briefHistoryNtwks}, which gives a high-level link between tables and networks that has been exploited in recent  algebraic statistics literature. Various problems relating to exact goodness-of-fit tests are well-studied for contingency table models and some ERGMs, for example, \cite{AHT2012}, \cite{DS98}, and  \cite{GPS16}.  
Such general  tests have not been developed for stochastic blockmodels and, in particular, the methodology does not directly apply to the models in which the node's block membership or the number of blocks are latent. 

In this paper, we derive finite-sample goodness-of-fit tests for three variants of the SBM. These variants include the classic SBM, which we call the ER-SBM; the degree-corrected SBM; and an intermediate model called the additive SBM. We present a Monte Carlo algorithm to implement the test, with desired levels of Type I and II errors. These tests are shown to work well even in the typical case of observing only one, possibly sparse network of arbitrary size, even where the effective sample size is small.  

One general goodness-of-fit test of an ER-SBM is provided in  \cite{lei2016goodness}. It is an asymptotic test, valid under certain  assumptions on relatively balanced community sizes as the number of nodes grows, and it is not applicable to the degree-corrected SBM.
While one can make an argument of validity of asymptotic tests in  simple model variants, finite-sample tests are more appropriate for more general models whose number of parameters grow linearly with $n$, such as the degree-corrected SBM. 
Of particular note is the fact that the testing method we derive is applicable to \emph{general} latent-block SBMs and, in fact, any model that is a \emph{mixture of log-linear models}.

\paragraph{Outline of the paper.}The three stochastic blockmodel  variants that we consider are introduced in \S~\ref{sec:models}. 
In \S~\ref{sec:ExactMethodology}, we describe frequentist and Bayesian conditional goodness-of-fit tests, along with the choice of a goodness-of-fit statistic, for both observed and latent block assignment of nodes; the latter is built from the former.
\S~\ref{sec:SBMExactTest} presents algebraic Monte Carlo algorithms for estimating the exact conditional $p$-value that are applicable to any variant of the SBM. These algorithms use some recently developed tools from algebraic statistics with the key ingredient being a description of so called \emph{Markov bases} that provides
a good way to sample from the conditional distribution given the sufficient statistics. \S~\ref{sec:Alg+Geom} derives the description of Markov bases for the three variants of the SBMs. This section also contains key theoretical results on the geometry of the three model polytopes with direct implications on the existence of MLE.
\S~\ref{sec:PowerAndComparison} presents simulations of the power of our test and compares our method to that of \cite{lei2016goodness}. 
In \S~\ref{sec:simulations} we assess the performance of our tests on synthetic graphs, on the well-known Karate dataset \cite{zachary1977information}, and the Human Connectome data (\url{http://www.humanconnectomeproject.org/}). 
We close with discussion in \S~\ref{sec:discussion}. The  proofs of the theoretical results appear in the Appendix, \cite{mrc2016-supplement}.

\section{Three variants of the stochastic blockmodel}
\label{sec:models}
Let $G$ be a random graph on $n$ nodes, $g$ being its realization. 
We assume that all graphs are unweighted and undirected, and that self-loops are not allowed.  We may refer to a graph by its $n \times n$  adjacency matrix, also denoted by $g$, in which $g_{uv} = 1$ if there is an edge between nodes $u$ and $v$ in the graph $g$, and $0$ otherwise. 

An exponential family random graph model (ERGM) assumes that the probability of observing a graph $g$ depends only on a vector of sufficient statistics denoted by $T(g)$. Formally,
the probability of observing a given network $G = g$ takes the exponential form  $$\P_{\theta}(G = g) = \frac{\exp\langle T (g) , \theta \rangle }{\psi(\theta)},$$ where $\psi(\theta) = \sum_g \exp \{\langle T (g) , \theta \rangle \}$ is   the normalizing constant, 
$\theta \in \Theta$ is a vector of natural parameters, and $T (g)$ is the vector of minimal sufficient statistics for the model.  
As is customary in the ERGM setting \citep{Robins2007}, the random graph $G$ is viewed as a collection of random variables, one for each pair of nodes. A pair of nodes is called a \emph{dyad}. When the graph is undirected, a dyad $\{i,j\}$ can be in only one of two states: 0 or 1, indicating absence or presence of an edge between nodes $i$ and $j$.  The term `dyad' will be used only when we intend to focus on the specific node-pair as a random variable. 
For undirected simple graphs, which are the focus of this paper,  the phrase `probability of occurrence of an edge between nodes $i$ and $j$' and the phrase `probability of the dyad $\{i,j\}$ being in state $1$' are interchangeable.  

Let $p_{uv}$ denote the probability of an edge between node $u$ and node $v$, where we assume $g_{uv} \sim$ Bernoulli$(p_{uv})$. A stochastic blockmodel (SBM) postulates that the nodes are partitioned into $k$ blocks and the probability $p_{uv}$  
depends on their block membership. 
Below, we define a generic version of a stochastic blockmodel with known and latent block assignments. Then, we consider three different \emph{log-linear} parametrizations of $p_{uv}$ which give rise to three variants of the SBM. 
When the block assignments are known, each variant, due to its log-linear parametrization, has an exponential family form familiar to the statistics literature. 
The log-linear viewpoint  allows us to formulate sufficient statistics of an SBM with known blocks. 
Sufficient statistics---crucial for developing the goodness-of-fit tests---and the corresponding model polytopes 
for each of the three model variants
are studied in detail in Section \ref{sec:Alg+Geom}.

\paragraph{A generic SBM with known and latent block assignments.} 
Assume that the nodes $[n] = \{1, \ldots, n\}$ can be partitioned into $k$ blocks $B_1, \ldots, B_k$. For ease of notation, block $B_i$ may be referred to simply by its label, $i$. 
The block assignment function $\mathcal Z:[n]\to [k]$ records the block label of each node; thus the block assignment is a list $\blocks =\{z(1),\dots,z(n)\}$, where each $z(u) \in \{1, \ldots k\}$. Given a block assignment $\mathcal{Z}$, and a matrix of edge probabilities $[p_{uv}]$, the probability of observing a graph $g$ takes the form
\begin{equation}\label{eqn general SBM}
	\P_{\theta}(G=g|\mathcal{Z} = z)  = \prod_{1\leq u,v\leq n} p_{uv}^{g_{uv}} (1-p_{uv})^{(1-g_{uv})},
\end{equation}
where  $p_{uv}$ depends on the known block assignment $\mathcal Z$.
The specific parametrization of each $p_{uv}$ is further governed by the model variant. 

In the Bayesian setting, when $\blocks$ is unknown, we treat it as a  random variable, and therefore  the probability of observing a graph $g$ takes the form
\begin{equation}\label{eqn general latent SBM}
	\P_{\theta}(G=g)  = \sum_{\mathcal{Z} = z} \prod_{1\leq u,v\leq n} p_{uv}^{g_{uv}} (1-p_{uv})^{(1-g_{uv})} \cdot \P(\mathcal{Z} = z).
\end{equation}
It is common to specify $\P(\mathcal{Z})$ by a multinomial distribution over $\{n_i\}$, where $n_i = |B_i|$, the size of block $B_i$. We discuss the specifics in the simulation sections that follow. 

\subsection{The Erd\"os-R\'enyi-SBM} In the classical version of the stochastic blockmodel from \cite{holland1983stochastic}, the  probability of an edge occurring between two nodes depends \emph{only} on their block assignments. To distinguish from other blockmodels below,  we will refer to this model as the Erd\"os-R\'enyi SBM, or ER-SBM. This model is a natural generalization of the classic Erd\"os-R\'enyi model. 

\begin{defn}[ER-SBM]\label{def:BlockModel}
	The ER-SBM is specified using ${k+1 \choose 2}$
	parameters, 
	$\alpha_{z(u)z(v)}$ for $1\leq u,v\leq n$, with log-odds of the probability of the edge $\{u,v\}$ given by: 
	\begin{eqnarray}\label{erSBM}
		\log \left(\frac{p_{uv}}{1-p_{uv}}\right)=\alpha_{z(u)z(v)}.
	\end{eqnarray}
\end{defn}
Let  nodes $u$ and $v$ belong to blocks $i$ and $j$, respectively, so that  $z(u) = i$ and $z(v) = j$.
The model parameter $\alpha_{ij}$ measures propensities of nodes in pairs of blocks $(i,j)$ to be connected when $i \neq j$, and the edge density within each block when $i=j$. 

Following standard notation, we can also define a $k\times k$ matrix $Q=[q_{ij}]$ where $q_{ij}$ denotes the probability of an edge between a node in block $B_i$ and a node in block $B_j$. Note that $q_{ij} = \exp(\alpha_{ij})/(1 + \exp(\alpha_{ij}))$. 
In this case, the probability  of observing a graph $g$ takes the following form:
\[
\P_{\theta}(G=g| \mathcal{Z})  = \prod_{u,v} q_{z(u)z(v)}^{g_{uv}} (1-q_{z(u)z(v)})^{(1-g_{uv})}.
\]

When $\mathcal Z$ is known, the ER-SBM can be written as an exponential family of the form $\P_{\theta}(G=g|\blocks = z)\propto \exp\langle T_{ER}(g) , \theta \rangle$  detailed in \cite[Equation (33)]{holland1983stochastic}.  
In this case, the natural parameter vector $\theta$ is given by the upper diagonal of the $k \times k$ symmetric matrix $[\alpha_{ij}]$ of logits of the edge probabilities between and within blocks. The natural parameter space is $\Theta =  \mathbb R^{{k+1 \choose 2}}$. The sufficient statistics  $T_{ER}(g)$ are given by the upper diagonal elements of the symmetric $k \times k$ matrix, where the $ij^{th}$ entry of the matrix counts the number of edges between nodes in block $B_i$ and $B_j$. Section \ref{sec:alg:BlockModel} offers a detailed study of this model.

\subsection{The Additive SBM}  The \emph{additive stochastic blockmodel} was introduced in \cite{fienberg1981categorical} and \cite{fienberg1985statistical} in the context of directed networks. 
The additive SBM is a special case of ER-SBM, in which the parameters  $\alpha_{ij}$  are assumed to be additive: $\alpha_{ij} = \alpha_i + \alpha_j$. 
\begin{defn}[Additive SBM]  \label{def:additiveSBM}
	The additive stochastic blockmodel is parameterized by $k$ block parameters, $\alpha_i$ for $i\in [k]$, where the log-odds of each edge is specified as 
	\begin{eqnarray}\label{addSBM}
		\log \left(\frac{p_{uv}}{1-p_{uv}}\right)  =\alpha_{z(u)} + \alpha_{z(v)} = \alpha_i + \alpha_j.
	\end{eqnarray}
\end{defn}

Similarly to the ER-SBM, when $\mathcal Z$ is known, the additive SBM can be written in the exponential family form with the natural parameter vector $\theta  = (\alpha_1, \ldots \alpha_k) $. The natural parameter space is given by $\Theta = \mathbb R^k$. The vector of sufficient statistics is given by $T_{Add}(g) = (x_1, \ldots x_k)$, where $x_i$ is the sum of degrees of nodes in Block $B_i$. See \S~\ref{sec:alg:additiveSBM} for additional details.

\begin{remark} 
	\label{rmk: additive sbm} 
	Additive SBM is a special case of the $p_1$ distribution, the stochastic blockmodel with reciprocity. 
	As such, the model is related---in two ways---to the $\beta$-model for random graphs  from \cite{CDS11}. The $\beta$-model is an ERGM with the degree sequence of the graph as its sufficient statistic. In particular, let $\beta = \{\beta_1, \ldots, \beta_n\}$ be a vector of parameters, the $\beta$-model assigns a probability proportional to $\exp\{\sum_{u=1}^n \beta_u d_u\}$ to a graph $g$ with degree sequence $\{d_1, \ldots, d_n\}$. 
	The additive SBM can be seen as a special case 
	with the additional constraint $\beta_u=\beta_v$ if $z(u)=z(v)$. 
	It can be seen also as its generalization: identifying all nodes in a block turns the additive SBM into a generalized $\beta$-model for multigraphs in which we allow self-loops, from \cite{RPF:11}. 
\end{remark}

\subsection{The $\beta$-SBM} One drawback of the simple SBM variants above is that they do not provide an adequate fit to networks where nodes have highly varying degrees within blocks. To model networks with heterogeneous degrees, \cite{karrer2011stochastic} propose the degree-corrected stochastic blockmodels.  
Here we focus on the  exponential family version of this model, denoted by $\beta$-SBM.
\begin{defn}[$\beta$-SBM]\label{def:betaSBM} 
	This log-linear model is parameterized by $k+1\choose 2$ 
	block parameters $\alpha_{z(u) z(v)}$ and $n$ node-specific parameters $\beta_u$ for $1\leq u\leq n$, with the log-odds of the probability of an edge $\{u,v\}$ given by:
	\begin{eqnarray}\label{betaSBM}
		\log \left(\frac{p_{uv}}{1-p_{uv}}\right) = \alpha_{z(u)z(v)} + \beta_u + \beta_v.
	\end{eqnarray}
\end{defn}
The $\beta$-SBM was formally introduced in \cite{Xiaolin2015thesis} as an exponential family  model equivalent to the network-literature-standard non-ERGM definition, under the name  `exponential-family degree-corrected SBM' or EXPDCSBM. We will instead refer to it as the $\beta$-SBM: the model  is a natural combination of the $\beta$-model and the classical SBM. In fact, it is a generalization of $\beta$-model   where probabilities of edges appearing are influenced not only by the   endpoints, but also their block membership, via additional parameters $\alpha_{z(u)z(v)}$. 

When $\mathcal Z$ is known, the $\beta$-SBM is an exponential family model with natural parameter vector $\theta = (\beta, \alpha)$ where $\beta = (\beta_1, \ldots, \beta_n)$ and $\alpha$ is the vector of the upper diagonal elements of the $k \times k$ matrix  $[\alpha_{i,j}]$. The natural parameter space is $\Theta = \mathbb R^{n} \times \mathbb R^{ {k+1}\choose 2}$. The vector of sufficient statistics $T_{\beta}$ contains the degree of each node $i\in[n]$ and the number of edges between each pair of blocks $B_i$ and $B_j$ for $1\leq i \leq j\leq k$.  Further details on   model structure are in \S~\ref{sec:alg:BetaSBM}.

\begin{remark}
	\label{remark:dcparam}
	In the network literature, the degree corrected SBM is parameterized as $p_{uv} = w_u w_v p_{z_{u} z_{v}}$ where $w_u \in (0,1)$, for all  nodes $u$, are node specific parameters. We prefer the log-linear parametrization since it allows us to write the $\beta$-SBM in an exponential family form, and one can recover the classic ER-SBM by setting $\beta_u = 0$ for all $u$, making the ER-SBM and $\beta$-SBM formally nested models. 	
\end{remark}

\section{Testing goodness-of-fit of the SBMs}
\label{sec:ExactMethodology}

Given that the model structure significantly changes depending on whether the block assignments are known or latent, one does not expect there to exist a   single generic approach to testing model fit of the SBMs. In  \S~\ref{sec:testTheory}, we take the frequentist approach to construct an exact conditional test for the known-block version of the SBMs. In \S~\ref{sec:testTheoryBayes}, we take the Bayesian approach to testing the model fit of latent-block SBMs, and describe how it is the natural extension of the known-block test. 
We define a  conditional $p$-value and derive two equivalent interpretations of it, which allows us to use the known-block test to implement the latent-block test. This is the key concept that allows us to use the algebraic statistics machinery for model testing, which is described in the next section, in case of models with latent variables for the first time in the literature.  Choices of goodness-of-fit statistics that are required for implementing all of these tests are discussed in \S~\ref{sec:GoFstatistic}. 

\subsection{A frequentist conditional goodness-of-fit test}
\label{sec:testTheory}
Consider any variant of the stochastic blockmodel $\P_{\theta}(G| \mathcal{Z})$ with a generic parameter vector $\theta$ and block assignment $\mathcal{Z}$. We will assume that the number of blocks $k$ is fixed and known. Consider the following goodness-of-fit test:
$$H_0: G \sim \P_{\theta}(G| \mathcal{Z})$$
against the general alternative. 
This is a composite null hypothesis with unknown parameters $\theta$ and $\mathcal{Z}$. When the block assignment $\mathcal{Z}$ is known,  $\P_{\theta}(G| \mathcal{Z})$ is an exponential family, and thus a natural conditional test for $H_0$ is to condition on the sufficient statistics. Specifically, let $T_{\mathcal{Z}}(g)$ denote the sufficient statistics when the block assignment is $\mathcal{Z}$. Let $\mathcal{Z}_0$ denote the true block assignment. The conditional distribution
$\P_{\theta}(G |\mathcal{Z}_0, T_{\mathcal{Z}_0}(g))$
is free of the parameters $\theta$ and hence can be used for performing a goodness-of-fit test. 

\begin{defn}[A fiber for a discrete exponential family]\label{defn:fiber}
	Consider an exponential family model 
	$\P_{\theta}(G = g) = \exp\langle T (g) , \theta \rangle/(\psi(\theta))$, where  ${\theta}$ is a vector of parameters and ${T(g)}$ is a vector of sufficient statistics.
	Define the following subset of the sample space: $$\mathcal F_{u} := \{g : {T(g)} = {u} \}.$$ 
	This set is called \emph{the fiber} of $u$ under the given exponential family model. It is the support of the conditional distribution given the sufficient statistics. 
\end{defn}
\begin{remark}\label{rmk:fiber notation with no Z}
	For an SBM with known $\mathcal Z$, the sufficient statistics, and hence the corresponding fiber, depends on $\mathcal Z$. The reader should note that in Definition~\ref{defn:fiber}, we suppress the possible dependence of $T(g)$ (and hence the fiber $\mathcal  F_u$) on $\mathcal Z$. This is because our testing setup and results are applicable more generally to exponential family models, and not just the SBMs. 
\end{remark} 

\begin{prop}
	\label{prop:uniformDistribution}
	Consider an exponential family model as in Definition~\ref{defn:fiber}. 
	Let $u$ denote a fixed value of the sufficient statistics. 
	Then the conditional  distribution on the fiber is  $$\P_{\theta}(G=g|{T(g)} = {u}) = \frac{1}{|\mathcal F_{u}|} \mbox{ if } g \in \mathcal{F}_{u}, 0 \mbox{ otherwise}.$$
\end{prop}
That this conditional distribution is uniform is stated, for example, in \cite[p.365]{DS98}.  For completeness, we include a short proof.

\begin{proof}
		The proof proceeds by an elementary calculation of the conditional distribution:
		\begin{align*}
			\P_{\theta}(G=g| T(g) = u) &= \frac{\P_{\theta}(G=g,T(g) = u )}{\P(T(g) = u)}\\
			&= \frac{\P(G=g)}{\P(T(g) = u)} \text{ if } g \in  \mathcal{F}_{u}, 0 \text{ otherwise}\\
			&= \frac{\P(G=g)}{\sum_{g' \in \mathcal F(u)} \P(G=g')} \\
			&= \frac{\exp(\theta^t \cdot T(g)) }{\sum_{g' \in \mathcal F_{u}} \exp(\theta^t \cdot T(g')) }\\
			&= \frac{\exp(\theta^t \cdot u)}{ \sum_{g' \in \mathcal F_{u}} \exp(\theta^t \cdot u) }\\
			&= \frac{1}{|\mathcal F_{u}|}.
		\end{align*}	
\end{proof}

Applying Proposition \ref{prop:uniformDistribution} to a stochastic blockmodel with known block assignments $\blocks_0$, we may thereby suppress the subscript $\theta$ from the notation of the conditional distribution $\P_{\theta}(G|\blocks_0, T_{\blocks_0}(g))$ on the fiber  $\mathcal F_u$ of the observed value of the sufficient statistic $u={T_{\blocks_0}(g)}$. 

To carry out a formal goodness-of-fit test, let  $\gof(g)$ 
be any goodness-of-fit statistic that is a function of the graph $g$ and the block assignment function $\mathcal{Z}$, such that large values of $\gof(g)$ imply departures from the model;  examples are contained in \S\ref{sec:GoFstatistic}. 
Such a goodness-of-fit statistic is said to be \emph{valid}, because it can lead to a valid exact conditional $p$-value. 
The conditional distribution of this statistic on the fiber is the reference distribution of the literature-standard exact conditional test for exponential families. 
Namely, when block assignment is known, the exact conditional $p$-value is 
\begin{equation}\label{eqn frequentist exact conditional p-value}
	p(\blocks_0, g) = \P\left( \gofz(G) \geq \gofz(g) | \blocks_0, T_{\blocks_0}(g)\right).
\end{equation}

When block assignments are unknown, as is often the case in practice, a natural frequentist approach is to replace $\blocks_0$ by a consistent estimator $\hat \blocks$. As in the case of known $\blocks$, the conditional distribution $ \P(\cdot | \hat \blocks, T_{\hat \blocks}(g))$ does not depend on $\theta$, and is uniform over $\mathcal F_{T_{\hat{\blocks}}(g)}$. The following is the \emph{plug-in} p-value of the goodness-of-fit test: 
$$p(\hat \blocks, g) = \P\left( \mbox{GoF}_{\hat\blocks}(G) 
\geq \mbox{GoF}_{\hat\blocks}(g) 
| \hat \blocks, T_{\hat \blocks}(g)\right).$$
A consistent estimator $\hat{\blocks}$ converges in probability to $\blocks_0$, 
therefore   the $p$-value $p(\hat \blocks, g)$  is expected to be \emph{asymptotically valid}. Asymptotics of such tests are studied in a follow-up paper.

\subsection{A Bayesian conditional goodness-of-fit test for latent blocks}  
\label{sec:testTheoryBayes}
 In the case when block assignment is unknown,  the frequentist conditional test in \S~\ref{sec:testTheory} ignores the uncertainty in the estimation of $\blocks$. To better handle the case of unknown $\blocks$, we construct a Bayesian version of the test as follows, considering $\blocks$ to be a latent variable. 

An SBM with latent blocks can be written as a mixture of exponential family models with known blocks. The geometry of mixtures is explained in \cite{FHRZ07-MLEforLatent}, which explores the link between the geometric and statistical model properties and the implications on parameter estimation. 
For such mixture models, the minimal sufficient statistic is the entire observed graph $g$, 
and the conditional goodness-of-fit test should be based on $\P(G|g)$. This is equivalent to computing the $p$-value based on the posterior predictive distribution $\P(G|g)$. The $p$-value is given by 
\begin{align}\label{eq:composite p-value}
	p_{\mbox{bayes}}^1(g) = \P^{\P(G | g)}(\gof(G) \geq \gof(g)),
\end{align} 
where $\P^{m}(E)$ denotes that the probability of the event $E$ is calculated under the distribution $m$.

While this construction is natural, there is a better way to interpret the  $p$-value above, which we will show to be equivalent and which can be used to build an efficient sampler to carry out the test in \S~\ref{sec:algorithms}. 
Namely, the Bayesian conditional goodness-of-fit test can be seen as a simple extension of the frequentist test. As in the frequentist setting, for each fixed value of $\blocks$, we compute the conditional $p$-value by conditioning on the sufficient statistics given that $\blocks$, i.e. $T_{\blocks}(g)$. (Recall that this distribution does not depend on $\theta$). Next, we compute a posterior distribution over $\blocks$, given by $\P(\blocks | g)$, and average the frequentist conditional $p$-values with respect to this distribution. In particular, 
\begin{align}\label{eq:composite p-value}
	p_{\mbox{bayes}}^2(g) = \sum_{\blocks\in Z_{n,k}} p(\blocks , g) \P(\blocks | g),
\end{align}
where $p(\blocks, g)$ is the exact conditional frequentist $p$-value assuming that the block assignment is known and $Z_{n,k}$ is the set of all  possible block assignments of $n$ nodes into $k$ blocks. The proposition below shows that these two constructions are equivalent.
\begin{proposition} 
	\label{prop:eq}
	With definitions as above, 	$p_{\mbox{bayes}}^1(g) = p_{\mbox{bayes}}^2(g)$. 
\end{proposition}
	\begin{proof}
		We begin by showing $
		\P(G| g) = \sum_{\blocks} \P( G|\blocks, T_{\blocks}(g), g) \cdot  \P(Z|g)$ as follows.  
		\begin{align*}
			\P(G = g'|g) &= \sum_{\blocks\in Z_{n,k}}\sum_{T_{\blocks}(g')} \P(G = g'|\blocks, T_{\blocks}(g'), g) \cdot \P(\blocks|g) \cdot \P(T_{\blocks}(g') | g) \\ 
			&= \sum_{\blocks\in Z_{n,k}} \P(\blocks|g) \cdot \left(\sum_{T_{\blocks}(g')} \P(G=g'|\blocks, T_{\blocks}(g'), g)  \P(T_{\blocks}(g') | g) \right)\\ 
			&= \sum_{\blocks\in Z_{n,k}} \P(\blocks|g) \cdot  \P(G=g'|\blocks, T_{\blocks}(g'), g), 
		\end{align*} 
		where the last equality holds because for a given observed graph $g$ and fixed value of $\blocks = \blocks_0$ in the outer sum, the term in the inner sum,  $\P(T_{\blocks}(g') | g) = 1$ when $T_{\blocks}(g') = T_{\blocks_0}(g')$ and $0$ for all other values of $T_{\blocks}(g')$.  Next, note that
		\begin{align*}
			p_{\mbox{bayes}}^1(g) &= \P^{\P(G \mid g)}(\gof(G) \geq \gof(g)) \\
			&= \sum_{\blocks\in Z_{n,k}}   \P(\gof(G) \geq \gof(g) \mid \blocks, T_{\blocks}(G), g) P(Z|g)\\
			&= \sum_{\blocks\in Z_{n,k}} p(\blocks, g) \P(\blocks \mid g)
			= p_{\mbox{bayes}}^2(g).
		\end{align*}
\end{proof}
The second construction lends itself to an efficient way to sample from the conditional distribution $P(G|g)$. In particular, 
$\P(G|g)$ can be written as following identity: 
\begin{align}\label{eq:BreakingUpSamplingForLatent}
	\P(G\mid g) = \sum_{\blocks\in Z_{n,k}} \P(G\mid \mathcal{F}_{T_{\blocks}(g)}) \cdot \P(\blocks\mid g), 
\end{align}
where $\mathcal{F}_{T_{\blocks}(g)}$ is the fiber. We take advantage of equation \eqref{eq:BreakingUpSamplingForLatent} in  sampling algorithms for a Monte Carlo estimator of the Bayesian $p$-value. This is the theory behind Algorithm \ref{algo:latentSBMtestGeneral}. 

\begin{remark}
	The Bayesian version of our test can also be interpreted as sampling from the posterior predictive distribution. Posterior predictive checks for model validation, such as \cite{meng1994posterior,gelfand1998model,gelman1996posterior},  are  popular  Bayesian counterparts of $p$-values and are suited to latent variables models or models with abundance of nuisance parameters. 
	Sampling from the posterior predictive distribution is challenging in general and equation \eqref{eq:BreakingUpSamplingForLatent} provides an efficient way to gather the samples by introducing an intermediate step of sampling from the {\em conditional fiber} .  
	The posterior quantity analogous to $p$-value here is the posterior predictive-$p$-value,  which can be viewed as the posterior mean of the classical $p$-value. Despite some  controversy regarding issues with calibration, we obtained promising results in delivering accurate Type I and II errors.  
\end{remark}

\subsection{Choosing a goodness-of-fit statistic} 
\label{sec:GoFstatistic}
The remaining theoretical question is the derivation of a valid goodness-of-fit statistic for SBM variants from \S~\ref{sec:models}.  
From a log-linear model perspective, one natural choice is  Pearson's chi-square statistic, a scaled distance between the observed $g$ and the expected $g$ under the null model: 
\begin{align}\label{eq:gofgen}
	\sum_{1\leq u<v\leq n} \frac{(\hat g_{uv}-g_{uv})^2}{\hat g_{uv}}. 
\end{align}
Chi-square is a reasonable  choice intuitively, because, it incorporates the expected graph, and thus large values of it indicate departure from the null model.  
For general log-linear exponential family models for contingency tables, the chi-square is valid and leads to bona-fide exact conditional tests, as explained in \cite{PRF10, SteveAleMe-holland} and  \cite{DS98}. 
For network models, there are subtleties related to model specification and the corresponding contingency table representation of $g$; \cite{Fienberg2012:briefHistoryNtwks} summarizes some ERGM table representations,  studied in detail in \cite{GPS21+}. 
The advantages of the log-linear representations outweigh the technical issues that arise, but one remains: the structure of the model fibers of such large sparse tables may bring into question the validity of the statistic. The reader should note that this question is \emph{not} due to the fact that the asymptotics do not apply in large, sparse table settings;  the conditional test construction does not require the derivation of the asymptotic distribution of the statistic. Rather, the question is whether or not the chosen statistic has a non-degenerate distribution: if it is constant on the fiber, then the reference conditional distribution on the fiber used to compute the $p$-value in Equation~\eqref{eqn frequentist exact conditional p-value} is meaningless. 

The fact is that in some simplistic models---such as the ER-SBM---the  chi-square statistic from Equation~\eqref{eq:gofgen} computed only using the adjacency matrix $g$ is  constant on all fibers. 
Such degenerate cases require the derivation of other test statistics for SBMs.  The choices of $\gof(g)$ below are quite deliberate for two reasons. First, they generalize Equation~\eqref{eq:gofgen} which is valid in the non-degenerate case. 
Second, the  thesis \cite{MatejaPhD} studies how the network breaks down into so called \emph{atomic} parts, each of which contribute to the sufficient statistic computation. This allows then a formal derivation of what is therein referred to as the atomic GoF statistic, which captures exactly the same information as the chi-square in Fisher's exact test. The GoF statistics below turn out to be atomic, and so in this formal sense, they are the most canonical choices for the finite-sample tests based on model fibers. 

{\bf GoF statistic for the ER-SBM.}
Define $m_{ui}$ to be the number of neighbors vertex $u$ has in block $B_i$. Note that $m_{ui}=\sum_{v\in B_i}g_{uv}$. 
A straightforward calculation shows that the entry  $q_{ij}$ of the $k\times k$ matrix $Q$ is the probability of an edge between block $B_i$ and block $B_j$. Recall \cite[p.1928]{BickelMLEsbm} that the MLE of $q_{ij}$ is:  
\begin{align} \label{eq:empirical}
	\hat q_{ij} = 
	\begin{cases}
		\frac{\sum_{u\in B_i}\sum_{v\in B_j}g_{uv}}{n_i n_j}, \mbox{ if } i\neq j,\\  \\ 
		\frac{\sum_{u \in B_j}\sum_{v \in B_j}g_{uv}}{n_i(n_i-1)}, \mbox{ if } i=j. 
	\end{cases}
\end{align} 
Note that the value of $\hat q_{ij}$ is constant on any given fiber, and hence needs to be computed only once when performing the conditional test. 
We define the block-corrected chi-square statistic as follows: 
\begin{equation}\label{eqn:DegreeCorrectedChiSquared}
	\gof(g)  = \chi^2_{\mbox{Block-corrected}}(g,\blocks) := \sum_{u=1}^n\sum_{i=1}^k \frac{(m_{ui}-n_i\hat q_{z(u)i})^2}{n_i\hat q_{z(u)i}}.
\end{equation}
Since $m_{ui}$ counts the number of neighbors $u$ has in block $B_i$, under the ER-SBM, the expected value $E[m_{ui}]=n_i q_{z(u)i}$,  therefore large values of $\chi^2_{\mbox{Block-corrected}}(g,\blocks)$,  in which we have replaced $q_{z(u)i}$ with the MLE $\hat q_{z(u)i}$, indicate lack of fit.

{\bf GoF statistic for the $\beta$-SBM.} Since the usual chi-square statistic is not constant on the fibers of the $\beta$-SBM, there is no need to use the corrected version of it above. 
Then, analogous to 
the $\beta$-model \cite{PRF10}, we use the usual Pearson's chi-square statistic: 
\begin{align}\label{eqn:ChiSquared}
	\gof(g)  =\chi^2_{\beta\text{SBM}} (g,\blocks) =& \sum_{1\leq u<v\leq n} \frac{( g_{uv}-\hat g_{uv})^2}{\hat g_{uv}}\\
	\text{where }		\hat{g}_{uv} =& \frac{\exp(\hat{\alpha}_{z(u)z(v)}+\hat{\beta_u}+\hat{\beta_v})}{1+\exp(\hat{\alpha}_{z(u)z(v)}+\hat{\beta_u}+\hat{\beta_v})}, \nonumber
\end{align}
where $\hat\alpha$ and $\hat\beta$ are MLEs as above. 
By definition, large  values of $\chi^2_{\beta\text{SBM}}$ correspond to departure from the null.

\section{Estimating conditional $p$-values: algebraic statistics Monte Carlo}
\label{sec:algorithms}
\label{sec:SBMExactTest}
\paragraph{Known block assignment.} 
An  SBM with a {known} block assignment is an example of an ERGM with a particularly nice structure: that of a log-linear model, in which  sufficient statistics are calculated as a linear function of the data.  Log-linear network models are equivalent to log-linear models for 0/1 contingency tables, a connection that directly imports fast estimation algorithms such as iterative proportional fitting. From the point of view of networks, this line of work dates back to \cite{fienberg1981categorical}, with \cite{Fienberg2012:briefHistoryNtwks} offering a more recent historical overview.
The connection to categorical data models also allows the use of algebraic statistics for exact conditional tests introduced in the seminal paper \cite{DS98} and extended with much detail in the monograph \cite{AHT2012}.   An important note is that the $0/1$ constraint on the table entries--stemming from the simple graph restriction--is a sampling constraint, making some of the standard algebraic statistics machinery less practical than one would like. This particular issue has been theoretically solved in the algebraic statistics literature. 
\cite{GPS21+} and \cite{ARSIA} give an overview of what the algebraic statistics can effectively do for network models, while \cite{KP:AOAS} illustrate and discuss the scaling and impact of  exact testing on networks. 

In the exact conditional test setup of \S~\ref{sec:ExactMethodology}, the $p$-value of the observed network $g$ with sufficient statistic $T(g)=u$ is computed as a proportion of the graphs sampled from the fiber  $\mathcal F_u$ whose goodness-of-fit statistic value is at least as extreme as the observed. Equation~\eqref{eqn frequentist exact conditional p-value} defines this $p$-value, and Definition~\ref{defn:fiber}  the fiber; see also  Remark~\ref{rmk:fiber notation with no Z} on notation.
Enumerating the fiber is computationally intractable, thus one samples from it using a Markov chain Monte Carlo algorithm. The key theorem in algebraic statistics from \cite{DS98}, often called the Fundamental Theorem of Markov Bases, is that there \emph{always} exists a finite set of steps, or moves, that one can use to sample from the conditional distribution on the fiber. This set is called a \emph{Markov basis}, and is guaranteed to connect all fibers of a given log-linear exponential family model. For any log-linear model, the existence and finiteness of such  a set is guaranteed by a fundamental result from algebra called the Hilbert basis theorem.  Finally, a Markov basis is always computable in finite time; these results are not purely existential.  \cite{WhatIsMB} gives a detailed overview and an applied example of using Markov bases.  
For completeness we provide a formal definition. Note that requiring $T(b)=0$ below  ensures that adding a table $b$ to $g$, where the sum is performed entry-wise, does not change the values of the sufficient statistics. 
\begin{defn}[A Markov basis for a discrete exponential family] \label{defn:MB}
	Adopt the fiber notation from Definition~\ref{defn:fiber}. Let $\mathcal B$ be any set of tables of same format as $g$ such that $T(b)=0$ for all $b\in\mathcal B$. 
	The set $\mathcal{B}$ is said to \emph{connect} the fiber $\mathcal{F}_u$ if given any two graphs $g_1,g_2\in\mathcal{F}_u,$ there exist moves in $\mathcal{B}$ that allow one to move from $g_1$ to $g_2,$ visiting only graphs in $\mathcal{F}_{u}$:
	$$ 
	g_2 = g_1+\sum_{i=1}^l b_i \mbox{ such that } b_i\in\mathcal B,\mbox{ and } g_1+\sum_{i=1}^j b_i \in\mathcal F_u\mbox{ for all } j\leq l.
	$$
	If $\mathcal{B}$ connects $\mathcal{F}_{u}$ for every  possible value of the sufficient statistic $u$, then $\mathcal{B}$ is said to be a \emph{Markov basis} for the log-linear exponential family with sufficient statistic $T$. 
\end{defn} 
The most important consequence is that using Markov bases to sample from the fiber leads to a connected and irreducible Markov chain. 
In Algorithm~\ref{algo:knownSBMtest} below, the exact conditional $p$-value of $g$ with $T(g)=u$ is estimated by one such Markov chain Monte Carlo algorithm, namely the Metropolis-Hastings algorithm, where each execution of Step~\ref{step:sampleFromFiber} in the algorithm produces a new graph in the fiber $\mathcal F_u$  using one Markov basis step. 
Markov bases for the SBM model variants are derived in Sections~\ref{sec:alg:BlockModel}, \ref{sec:alg:additiveSBM}, and \ref{sec:alg:BetaSBM} using the algebraic correspondence. 
Since the bases are very large, we do not pre-compute them a priori, rather we construct one basis element at random following the dynamic  construction of \cite{GPS16}. The quantity $f_{pval}$ in Step~\ref{step:fraction of graphs in fiber} is the fraction of graphs in the fiber whose $\gof$ values are at least as extreme as the observed. 
The output $pval$ of  Algorithm~\ref{algo:knownSBMtest}, computed in Step~\ref{step:p value}, is a Monte Carlo estimate of the exact conditional $p$-value from Equation~\eqref{eqn frequentist exact conditional p-value}. 
The exact testing procedure described herein has not yet appeared in the SBM literature, although it uses the standard method for testing models on discrete data in algebraic statistics. It applies to all known-block SBMs from \S~\ref{sec:models}. The main motivation for recasting the Markov bases Monte Carlo algorithm specifically for SBMs  is that it serves as a  building block for latent SBMs.

{\scriptsize
	\begin{algorithm}[h]
		\LinesNumbered
		\DontPrintSemicolon
		\SetAlgoLined
		\SetKwInOut{Input}{input}
		\SetKwInOut{Output}{output}
		\Input{$g$, an observed graph on $n$ nodes,\\
			$T(\cdot{})$, choice of a sufficient statistic; equivalently, a choice of an SBM,\\ 
			$\blocks=\{z(1),\ldots,z(n)\}$, a fixed block assignment,\\
			$\gof(\cdot{})$, choice of goodness-of-fit statistic,\\
			$\gof(g)$, the observed value of the goodness-of-fit statistic,\\
			numGraphs, the number of graphs to sample from the fiber $\mathcal F_{T(g)}$. 
		}
		\Output{$p$-value for the hypothesis test that the chosen model fits $g$ against a general alternative, and the reference sampling distribution.}
		\BlankLine
		Set $g_0=g$, the initial point on the fiber to be the observed graph\;
		\For{each $i=1$ \KwTo $\mathrm{numGraphs}$}{
			Randomly construct a  move $m$ in a Markov basis connecting the fiber $\mathcal F_{T(g)}$  \label{step:randomMarkovMove} \;
			Use $m$ to construct  $g_i\in\mathcal{F}_{T(g)}$   from the current graph $g_{i-1}$ \label{step:sampleFromFiber}\;
			Compute $\gof(g_i)$  \label{step:GoFsOnFiber}
		}
		Compute $f_{\mathrm{pval}}:= 
		\#\{i : \gof(g_i)\geq \gof(g)\}$ \label{step:fraction of graphs in fiber}\\
		Let $pval=\frac{1}{\mathrm{numGraphs}}\cdot f_{\mathrm{pval}}$ \label{step:p value}\\
		Return the $pval$ and the sampling distribution  $\{\gof(g_i)\}_{i=0}^{numGraphs}$. \label{step:return sampling distribution}
		\caption{Goodness-of-fit test for an SBM with a known block assignment}
		\label{algo:knownSBMtest}
	\end{algorithm}
}

\label{sec:SBMExactTestLatent}
\paragraph{Latent block assignment.} \label{latent block assignment}
The  goodness-of-fit test for {latent-block SBMs} is constructed in Algorithm~\ref{algo:latentSBMtestGeneral}. To the best of our knowledge, this is the first non-asymptotic general goodness-of-fit test for latent SBMs in the literature. 
It is a two-phase algorithm: first it estimates the block assignment $\blocks$, and then calls  Algorithm~\ref{algo:knownSBMtest} using the estimate $\hat\blocks$ as input. 

In practice, there is some choice on how to  implement the estimation Step~\ref{step:estimateBlocks} of Algorithm~\ref{algo:latentSBMtestGeneral}. 
There is a growing body of work on algorithms for estimating the block assignment from stochastic block models; see e.g.,  \cite{amini2013pseudo,pati2015optimal,geng2016probabilistic,yan2016convex}.  Roughly speaking, they can be categorized into two large families, one generates a point estimator for each node, and the other generates the entire block assignment distribution.   
In a frequentist framework, replacing the posterior distribution of $ \blocks $ by a point estimate $\hat\blocks$ results in the valid $p$-value derived in \S~\ref{sec:testTheory}.  
In a Bayesian framework, assuming the number of blocks $k$ to be known, we use the method in \cite{pati2015optimal} to get a posterior distribution of the model parameters, $\P(\blocks \mid g,k)$, then sample block assignment from it.  The simulation studies in  \cite{pati2015optimal} demonstrate that the method leads to clustering consistency which ensures that the $\chi^2$-based test is asymptotically valid.  
When the number of blocks is unknown, one can use 
the mixture-of-finite-mixtures (MFM) method for SBM from  \cite{geng2016probabilistic}. This is a method that is provably consistent; 
see also \cite{NewmanReinert2016EstimatingNumBlocks} for another algorithm for which there is heuristic evidence  of  consistency. 
Another  Bayesian method for estimating configurations or block membership in the degree-corrected SBM is provided in  \cite{peng2013bayesian}. 
When the graph grows in size, the MCMC used to estimate $\blocks$ 
might suffer from slow convergence, and we can apply deterministic estimators instead. For example, in some of the simulations in this paper,  we use regularized spectral clustering \citep{amini2013pseudo}, which has been shown to be consistent asymptotically and achieve state-of-the-art performance. 
When the estimation is accurate, this delivers the same test statistics as when we know the ground-truth block assignments.

Algorithm~\ref{algo:latentSBMtestGeneral} returns the  Bayes $p$-value from Equation~\eqref{eq:composite p-value}. It takes advantage of Equation~\eqref{eq:BreakingUpSamplingForLatent} to sample from the distribution $\P(G|g)$ by first sampling a block assignment $\hat\blocks_{i}$ from $\pi = \P(\blocks|g)$ and then computing the $p$-value using the distribution $\P(G|\mathcal{F}_{T_{\blocks_{i}}(g)})$.   
In practice, to speed up the computation, the Bayes $p$-value is estimated by averaging the conditional $p$-value estimates from each fiber whose posterior probability was higher than a given threshold $\epsilon$ (Line~\ref{step:avgpValueComputation}). This threshold is computed on Line~\ref{step:compute threshold} and, roughly,  prunes away those block assignments which appear only once in the distribution $\pi$.

{\scriptsize
	\begin{algorithm}[h]
		\LinesNumbered
		\DontPrintSemicolon
		\SetAlgoLined
		\SetKwInOut{Input}{input}
		\SetKwInOut{Output}{output}
		\Input{
			$g$, an observed graph on $n$ nodes,\\ 
			$T(\cdot{})$, choice of a sufficient statistic; equivalently, a choice of an SBM,\\ 
			$\gof(\cdot{})$, choice of goodness-of-fit statistic,\\
			numGraphs, length of each fiber walk \\
		}
		\Output{$p$-value for the hypothesis test that the chosen model fits $g$ against a general alternative, and the reference sampling distribution(s).}
		\BlankLine
		Estimate a distribution, $\pi = \P(\blocks|g)$, of the block assignment $\blocks=\{z(1),\dots,z(n)\}$ given $g$\label{step:estimateBlocks} \;
		Set $\epsilon:= 1/m$,  where  $m$ is the support size of the estimated distribution $\pi$ \label{step:compute threshold} \; 
		Construct $\tilde\pi$ from $\pi$ by thresholding: $\tilde\pi:=\{\hat\blocks : \pi(\hat\blocks)>\epsilon \}$ \label{step:check threshold} \;
		Set $numFibers := |support(\hat\pi)|$, the number of distinct block  assignments appearing with significant probability\;
		\For{$j=1$ to $numFibers$}{
			Sample a block assignment $\hat\blocks_j$ 
			from the distribution $\hat\pi$ 
			\label{step:samplec}\;
			Compute $\gofzhatiterated(g)$ \; 
			Compute the $j$-th  value $pval_j$ and  sampling distribution  $\{\gof(g_i)\}_{i=0}^{numGraphs}$ by 
			Algorithm~\ref{algo:knownSBMtest} with the following inputs:  $g$, 	$T(\cdot{})$,  $\blocks=\hat\blocks_j$, 	$\gof(\cdot{})$, $\gofzhatiterated(g)$, numGraphs.  \label{step:call algorithm for known blocks}
		}
		Return $\sum_{j} \pi(\hat\blocks_{j}) \cdot pval_j$ and the corresponding sampling distributions   $\{\gofzhatiterated(g_i)\}_{i=0}^{numGraphs}$. \;   \label{step:avgpValueComputation} 
		\caption{
			A general goodness-of-fit test for an SBM with latent block assignment}
		\label{algo:latentSBMtestGeneral}
	\end{algorithm}
}

In the interest of space, we defer several illustrations of the algorithm to \S~\ref{sec:simulations}. 
For example,  Figure~\ref{fig:accept} depicts the output of Algorithm~\ref{algo:latentSBMtestGeneral} for a sparse graph on $n=90$ nodes partitioned into $k=2$ blocks of roughly the same size, with an unknown block assignment $\blocks$.  
After thresholding, there are three distinct  estimated block assignments $\hat\blocks_j$ such that $\pi(\hat\blocks_j)>1/2000$. 
The three histograms  correspond to the outputs of Algorithm~\ref{algo:knownSBMtest} for   block assignments $\hat\blocks_1, \hat\blocks_2, \hat\blocks_3$.  
The statistic $\gofzhatiterated(g)$ is the block-corrected chi-square $\chi^2_{\mbox{Block-corrected}}(g,\blocks)$ from Equation~\eqref{eqn:DegreeCorrectedChiSquared}; each of its  sampling distributions are  computed in Step~\ref{step:GoFsOnFiber}---and returned in Step~\ref{step:return sampling distribution}---of Algorithm~\ref{algo:knownSBMtest}.

\section{Algebra and Geometry of SBMs}  
\label{sec:Alg+Geom}
There are two key objects associated with each SBM, namely, model polytopes and Markov bases, whose structure drives fast estimation algorithms and the model tests presented herein. We summarize and illustrate the main results in this section, defering all proofs and technical discussion to the Appendix. 
In this section we discuss explicit maps to and from the sample space, thus we introduce the notation   $\mathbb{G}$ for  the sample space of all graphs on $n$ nodes. 
\begin{defn}
	Let $\M$ be a discrete exponential family model  with sufficient statistics $T$, and let $\T := \{T(g) \in \R^d : g\in\mathbb{G}\}$ be the range of sufficient statistics $T : \mathbb{G}\longrightarrow \R^d$, where $\mathbb{G}$ is the sample space. 
	The \emph{model polytope} is 
	$
	P_\M :=\conv(\T)\subset\R^d
	$, the convex hull of all observable sufficient statistics. 
\end{defn} 
The combinatorial geometry of $P_\M$ plays an important role in parameter estimation for the model $\M$. 
Specifically, by the standard theory of exponential families \citep{barndorff2014information}, the maximum likelihood estimator for the observed data exists if and only if the observed sufficient statistic vector lies in the relative interior of the model polytope $P_\M$.   
Therefore, it is worthwhile to have an explicit representation of $P_\M$ as an intersection of finitely many closed halfspaces in $\R^d$; that is,  an \emph{H-representation} of $P_\M$.  Furthermore, one  seeks  a representation with as few hyperplanes as possible. 
\cite{GeyerExpFamFan} identifies  the rays of the normal fan of this model polytope with `directions of recession' for the likelihood function of the model. 

Explicit descriptions of the model polytopes are not known for many models, one notable exception being the $\beta$-model \cite{RPF:10} and its variants (such as the Bradley-Terry model). 
Theorems~\ref{thm: ER-SBM polytope}, \ref{thm: additive SBM model polytope} and \ref{prop: beta-SBM model polytope} provide an $H$-representation for each SBM defined in \S~\ref{sec:models}. 
In addition, there are only polynomial-many hyperplanes in the parameter $k$, and as such efficient linear optimization techniques provide a fast test for MLE existence.  
The usefulness of such a representation goes beyond only the problem of MLE existence; for example,  it has important implications in data privacy and confidentiality and inference under noisy data  \cite{SesaVishesBetaPrivacy,SesaVishesBetaPrivacyAOS}.

Markov Bases, formally defined in Definition~\ref{defn:MB}, are necessary for the implementation of Step~\ref{step:randomMarkovMove} of Algorithm~\ref{algo:knownSBMtest}, and thereby also Step~\ref{step:call algorithm for known blocks} of Algorithm~\ref{algo:latentSBMtestGeneral}.  For a given network model, one could derive a set of moves that sample all graphs with the given sufficient statistic using graph theory, but proving that a proposed set of moves connects each fiber is not trivial. Rather than going through long combinatorial constructions, we derive Markov bases using algebra. Specifically, we describe the \emph{ideal of the model}, which is the set of all infinitely many polynomials which vanish on all points on the model. Given such a general description,  connectedness of all Markov basis chains is guaranteed by \cite[Theorem 3.1]{DS98}. In practice, it also allows us to modify the ideal bases---as a Markov basis is not unique---according to \emph{any} sampling constraints that may arise in applications; see \cite[Theorem 3.1]{GPS21+}. 

The algebraic constructions needed to derive the Markov bases are detailed in the Appendix; this section derives the Markov moves needed for implementation. Specifically, Theorems~\ref{graver er-sbm main section} and \ref{markov add-sbm main section} are corollaries of Theorems A.2 and B.2, respectively.  Example~\ref{ex: beta-SBM example} illustrates the Markov move construction for $\beta$-SBM, although the full algebraic description of the model ideal for the $\beta$-SBM is an open problem; cf. \S~\ref{sec:discussion}. 
Figures~\ref{fig: ER-SBM markov move}, \ref{fig: markov basis for additive SBM} and \ref{fig: markov move beta SBM}, along with simple explanations of how to interpret algebraic Markov moves in terms of the network data, illustrate the meaning and use of these results in practice.

\subsection{The ER-SBM}
\label{sec:alg:BlockModel}

The ER-SBM, Definition~\ref{def:BlockModel}, considers a partition of the node set $[n]$ into $k$ nonempty blocks $\B := \{B_1,\ldots,B_k\}$ and assigns a   parameter $\alpha_{z(u)z(v)}$ to the log-odds of the probability of  existence of the edge
$\{u,v\}$ in a graph $g$ that depends only on the blocks containing nodes $u$ and $v$.  
Consequently, the sufficient statistics $T_{\mathrm{ER}}$ for this model consist of a $k\times k$ symmetric matrix in which the $ij^{th}$ entry records the number of edges between the nodes in block $B_i$ and the nodes in block $B_j$.  
Equivalently, 
\begin{equation*}
	\begin{split}
		T_{\mathrm{ER}}&:\mathbb{G}\longrightarrow\R^{k+1\choose 2};\\
		T_{\mathrm{ER}}&:g\mapsto \left(T(g)_{ij} : 1\leq i \leq j \leq k\right),\\
	\end{split}
\end{equation*}
where 
$$
T_{\mathrm{ER}}(g)_{ij} := \left|\{\{u,v\}\in E(g) : u\in B_i,v\in B_j\}\right|.
$$
The proof of the following $H$-representation for the ER-SBM model polytope $P_{\M_{\mathrm{ER}}}$ can be found in Appendix A. 
\begin{theorem}
	\label{thm: ER-SBM polytope}
	Let $\M_{\mathrm{ER}}$ be an ER-SBM with node set $[n]$ partitioned into nonempty blocks $\B := \{B_1,\ldots,B_k\}$, and let $n_i:=|B_i|$ for all $i\in[k]$.  
	Then the model polytope $P_{\M_{\mathrm{ER}}}$ is the ${k+1\choose 2}$-dimensional box
	$$
	\prod_{i\neq j}\left[0,n_in_j\right]\times\prod_{i=1}^k\left[0,{n_i\choose 2}\right].
	$$
	In particular, $P_{\M_{\mathrm{ER}}}$ is the intersection of the closed halfspaces given by the set of inequalities
	\begin{equation*}
		\begin{split}
			0\leq &x_{ij} \leq n_in_j, \mbox{ for $1\leq i < j\leq k$, and}\\
			0\leq &x_{ii} \leq {n_i\choose 2}, \mbox{ for $i\in[k]$}.\\
		\end{split}
	\end{equation*}
\end{theorem}

Note that the size of this representation is polynomial in $k$, the number of blocks. 
\begin{remark}
	\label{rmk: ER-SBM MLE existence criterion}
	It is a fact from standard exponential family theory that the MLE exists if and only if the vector of sufficient statistics belongs to the relative interior of  $P_{\M_{\mathrm{ER}}},$ see, for e.g., \citep{barndorff2014information, brown1986, lauritzen1996graphical}. Given this fact, Theorem~\ref{thm: ER-SBM polytope} implies that the maximum likelihood estimator for any observed network under an ER-SBM cannot exist if any of the blocks are of size one. This is because,  if the observed network is of size $1$, the sufficient statistics necessarily correspond to a lattice point. If any of the blocks are of size one, there are simply no lattice points in the relative interior of $P_{\M_{\mathrm{ER}}}.$ 
\end{remark}

Recall Definition~\ref{defn:MB} of a Markov basis  as a set of moves that connects the space of all graphs with a fixed value of the sufficient statistic.  
\begin{theorem}[Corollary of Theorem A.2]\label{graver er-sbm main section}
	Let $\M_{\mathrm{ER}}$ be an ER-SBM. 
	A  Markov basis for $\M_{\mathrm{ER}}$ consists of replacing any edge between blocks $B_i$ and $B_j$ for $i,j\in[k]$ with any other edge connecting the same blocks.  
\end{theorem}
The proof of this result uses algebraic techniques to derive the full description of the ideal of the model and a strategy that generalizes to more complicated models. 
While technical details are relegated to the appendix, we offer a couple of vignettes for the reader interested in how the method works. 
To derive the  basis, we identify a finite collection of binomials generating the toric ideal associated with the ER-SBM.  
Each binomial can be recovered from the model parametrization and corresponds to one Markov move. The binomials  are expressed  in the indeterminates $x_{uv}$  representing the value $\frac{p_{uv}}{1-p_{uv}}$. 
Since the indeterminate $x_{uv}$ corresponds to the odds of the probabilities $p_{uv}$ of the edge $\{u,v\}$ appearing in the graph, a binomial such as $x_{uv}-x_{yw}$ for $u,y\in B_i$ and $v,w\in B_j$ corresponds to a move from a graph $g$ containing the edge $\{u,v\}$ but not the edge $\{y,w\}$ to a graph $h$ containing the edge $\{y,w\}$ but not the edge $\{u,v\}$.  
An example of one such move is depicted in Figure~\ref{fig: ER-SBM markov move}.
These moves are precisely the moves captured by Theorem A.2.
\begin{figure}
	\begin{center}
		\begin{subfigure}{.3\textwidth}
			\begin{tikzpicture}
				\draw [black, line width=.55mm] (0,.5) rectangle (1,2.5) node [black, above=9, left=2] {$B_1$};
				\draw [black, line width=.55mm] (5,0) rectangle (4,3) node [black, above=8, right=4] {$B_2$};
				\draw [black, line width=.55mm] (2,-1.5) rectangle (3,-.5) node [black, below=37, left=4] {$B_3$};
				\draw [black,fill] (.6,1.95) circle [radius=0.079] node [black,left=2] {1};
				\draw [black,fill] (.6,1.05) circle [radius=0.079] node [black,left=2] {2};
				\draw [black,fill] (4.4,2.35) circle [radius=0.079] node [black,right=2] {3};
				\draw [black,fill] (4.4,1.45) circle [radius=0.079] node [black,right=2] {4};
				\draw [black,fill] (4.4,.55) circle [radius=0.079] node [black,right=2] {5};
				\draw [black,fill] (2.5,-.9) circle [radius=0.079] node [black,below=2] {6};
				\draw [black] (.6,1.05)
				to (4.4,.55);
				\draw [black] (4.4,2.35)
				to (2.5,-.9);
				\draw [black] (.6,1.95)
				to (4.4,.55);
				\draw [black] (.6,1.95)
				to (4.4,2.35);
				\draw [black] (.6,1.05)
				to (4.4,1.45);
				\draw [black, line width=.0003mm] (4.4,2.35)
				to (4.4,1.45);
				\draw [black] (2.5,-.9)
				to (4.4,.55);
			\end{tikzpicture}
			\caption{A graph $g$ with blocks $\mathbb{B}$ and $T_{\mathrm{ER}}(g)=(0,4,0,1,2,0).$ }
			\label{fig:betaSBMg1}
		\end{subfigure}
		\ \ \ \ \ 
		\begin{subfigure}{.3\textwidth}
			\begin{tikzpicture}
				\draw [black, line width=.55mm] (0,.5) rectangle (1,2.5) node [black, above=9, left=2] {$B_1$};
				\draw [black, line width=.55mm] (5,0) rectangle (4,3) node [black, above=8, right=4] {$B_2$};
				\draw [black, line width=.55mm] (2,-1.5) rectangle (3,-.5) node [black, below=37, left=4] {$B_3$};
				\draw [black,fill] (.6,1.95) circle [radius=0.079] node [black,left=2] {1};
				\draw [black,fill] (.6,1.05) circle [radius=0.079] node [black,left=2] {2};
				\draw [black,fill] (4.4,2.35) circle [radius=0.079] node [black,right=2] {3};
				\draw [black,fill] (4.4,1.45) circle [radius=0.079] node [black,right=2] {4};
				\draw [black,fill] (4.4,.55) circle [radius=0.079] node [black,right=2] {5};
				\draw [black,fill] (2.5,-.9) circle [radius=0.079] node [black,below=2] {6};
				\draw [blue, loosely dashed, thick] (.6,1.05)
				to (4.4,2.35);
				\draw [red, densely dashed, thick] (.6,1.95)
				to (4.4,.55);
			\end{tikzpicture}
			\caption{A linear Markov move for the ER-SBM with blocks $\mathbb{B}$.}
		\end{subfigure}
		\ \ \ \ \ 
		\begin{subfigure}{.3\textwidth}
			\begin{tikzpicture}
				\draw [black, line width=.55mm] (0,.5) rectangle (1,2.5) node [black, above=9, left=2] {$B_1$};
				\draw [black, line width=.55mm] (5,0) rectangle (4,3) node [black, above=8, right=4] {$B_2$};
				\draw [black, line width=.55mm] (2,-1.5) rectangle (3,-.5) node [black, below=37, left=4] {$B_3$};
				\draw [black,fill] (.6,1.95) circle [radius=0.079] node [black,left=2] {1};
				\draw [black,fill] (.6,1.05) circle [radius=0.079] node [black,left=2] {2};
				\draw [black,fill] (4.4,2.35) circle [radius=0.079] node [black,right=2] {3};
				\draw [black,fill] (4.4,1.45) circle [radius=0.079] node [black,right=2] {4};
				\draw [black,fill] (4.4,.55) circle [radius=0.079] node [black,right=2] {5};
				\draw [black,fill] (2.5,-.9) circle [radius=0.079] node [black,below=2] {6};
				\draw [black] (.6,1.05)
				to (4.4,.55);
				\draw [black] (4.4,2.35)
				to (2.5,-.9);
				\draw [black] (.6,1.05)
				to (4.4,1.45);
				\draw [black] (.6,1.95)
				to (4.4,2.35);
				\draw [black] (.6,1.05)
				to (4.4,2.35);
				\draw [black, line width=.0003mm] (4.4,2.35)
				to (4.4,1.45);
				\draw [black] (2.5,-.9)
				to (4.4,.55);
			\end{tikzpicture}
			\caption{A graph $h$ with $T_{\mathrm{ER}}(h)=T_{ER}(g)$. }
			\label{fig:betaSBMg1}
		\end{subfigure}
		\caption{Two graphs $g$ and $h$ in the same fiber of the ER-SBM with block structure $\mathbb{B}=\{B_1, B_2, B_3\}$ and a Markov move corresponding to the linear form $x_{23}-x_{15}\in\ker(\varphi_{\mathrm{ER}})$ that moves from $g$ to $h$.  The blue, loosely dashed line indicates edge insertion and the red, densely dashed line indicates edge deletion.}
		\label{fig: ER-SBM markov move}
	\end{center}
\end{figure}


\subsection{The additive SBM} 
\label{sec:alg:additiveSBM}

The additive SBM, Definition~\ref{def:additiveSBM}, considers a partition of the node set $[n]$ into $k$ nonempty blocks $\B := \{B_1,\ldots,B_k\}$.  
As before, we set $n_i := |B_i|$ for all $i\in[k]$.
The sufficient statistics map for the additive SBM model is: 
\begin{equation*}
	\begin{split}
		T_{\mathrm{Add}}&:\mathbb{G}\longrightarrow\R^k;\\
		T_{\mathrm{Add}}&:g\mapsto (x_1\ldots,x_k),\\
	\end{split}
\end{equation*}
where 
$$
x_i := \sum_{v\in B_i}\deg_g(v).
$$
Recall from Remark~\ref{rmk: additive sbm} that the additive SBM is a generalization of the $\beta$-model on $k$ nodes that allows for multiple edges and loops.  
The model polytope of the typical $\beta$-model is the polytope of degree sequences for simple graphs, and the vertices of this polytope are well-known to correspond to degree sequences of \emph{threshold graphs} \citep{MP95}.  
Thus, one would hope that the geometry of the additive SBM model polytope as compared to that of the typical $\beta$-model reflects the underlying generalization.  
Indeed, the following theorem nicely captures this relationship.  
\begin{theorem}
	\label{thm: additive SBM model polytope}
	Let $\M_{\mathrm{Add}}$ be an additive SBM model defined on node set $[n]$ with block structure $\B := \{B_1,\ldots,B_k\}$.
	Let $T,S\subset\B$ be such that 
	\begin{equation}
		\label{eqn: subset conditions}
		T\cup S \neq \emptyset 
		\qquad
		\mbox{and}
		\qquad
		T\cap S = \emptyset.  
	\end{equation}
	The model polytope $P_{\M_{\mathrm{Add}}}$ is the intersection of the closed halfspaces defined by the inequalities
	\begin{equation}
		\label{eqn: additive sbm inequalities}
		\sum_{B_i\in T} x_i - \sum_{B_i\in S} x_i 
		\leq 
		2{\sum_{B_i\in T}n_i\choose 2} + \sum_{ \substack{B_j \in \mathbb{B} \setminus (T\cup S) \\ B_i \in T} }n_in_j,
	\end{equation}
	ranging over all such pairs of subsets of $\B$.  
\end{theorem}
The proof of Theorem~\ref{thm:  additive SBM model polytope} can be found in Appendix B.
Similar to Theorem~\ref{thm: ER-SBM polytope}, this representation is also polynomial in $k$, making it useful for efficiently assessing the existence of MLE.

To derive a Markov basis for the additive model, one considers the description of the minimal sufficient statistics given at the beginning of the section. It  is easy to see that the following set of moves indeed lies in the Markov basis, but to prove that they are enough to connect the fiber, i.e. reach all possible graphs with the same value of the sufficient statistics, is non-trivial. But the use of appropriate algebraic techniques, based on a famous and well-investigated set of binomials,  streamlines the proof. 

\begin{theorem}[Corollary of Theorem B.2] \label{markov add-sbm main section}
	\label{thm: corollary to markov basis theorem for additive SBM}
	A Markov basis for $\M_{\mathrm{Add}}$ is given by an exchange of edges along any sorted $4$-cycle and the exchange of any edge for any other edge whose endpoints lie in the same respective blocks.  
\end{theorem}

In this Markov basis, we now have the same moves as the ER-SBM, but also a collection of ``quadratic" moves corresponding to switching edges along a $4$-cycle.    
For example, the binomial $x_{uv}x_{yw}-x_{uy}x_{vw}$ for $u,w\in B_i$ and $v,y\in B_j$ corresponds to a move from a graph $g$ containing the two edges $\{u,v\}$ and $\{y,w\}$ but not the edges $\{u,y\}$ and $\{v,w\}$ to a graph $h$ containing the edges $\{u,y\}$ and $\{v,w\}$ but not the edges $\{u,v\}$ and $\{y,w\}$.  
Collectively, the four edges $\{u,v\}, \{y,w\}, \{u,y\}$ and $\{v,w\}$ form a $4$-cycle.  
An example of one such move is depicted in Figure~\ref{fig: markov basis for additive SBM}.
\begin{figure}
	\begin{center}
		\begin{subfigure}{.3\textwidth}
			\begin{tikzpicture}
				\draw [black, line width=.55mm] (0,.5) rectangle (1,2.5) node [black, above=9, left=2] {$B_1$};
				\draw [black, line width=.55mm] (5,0) rectangle (4,3) node [black, above=8, right=4] {$B_2$};
				\draw [black, line width=.55mm] (2,-1.5) rectangle (3,-.5) node [black, below=37, left=4] {$B_3$};
				\draw [black,fill] (.6,1.95) circle [radius=0.079] node [black,left=2] {1};
				\draw [black,fill] (.6,1.05) circle [radius=0.079] node [black,left=2] {2};
				\draw [black,fill] (4.4,2.35) circle [radius=0.079] node [black,right=2] {3};
				\draw [black,fill] (4.4,1.45) circle [radius=0.079] node [black,right=2] {4};
				\draw [black,fill] (4.4,.55) circle [radius=0.079] node [black,right=2] {5};
				\draw [black,fill] (2.5,-.9) circle [radius=0.079] node [black,below=2] {6};
				\draw [black] (.6,1.05)
				to (4.4,.55);
				\draw [black] (4.4,2.35)
				to (2.5,-.9);
				\draw [black] (.6,1.95)
				to (4.4,.55);
				\draw [black] (.6,1.95)
				to (4.4,2.35);
				\draw [black] (.6,1.05)
				to (4.4,1.45);
				\draw [black, line width=.0003mm] (4.4,2.35)
				to (4.4,1.45);
				\draw [black] (2.5,-.9)
				to (4.4,.55);
			\end{tikzpicture}
			\caption{A graph $g$ with blocks $\mathbb{B}$ and $T_{\mathrm{Add}}(g)=(4,8,2).$ \ \ \ \ \ \ \ \ \ \ \ \ \ \ \ \ \ \ \ \ \ \ \ \ \ \ \ \ \ \ \ \ \ \ \ \ \ \ \ \ \ \ \ \ \ \ \ \ \ \ \ \ \ \ \ \ }
			\label{fig:betaSBMg1}
		\end{subfigure}
		\ \ \ \ \ 
		\begin{subfigure}{.3\textwidth}
			\begin{tikzpicture}
				\draw [black, line width=.55mm] (0,.5) rectangle (1,2.5) node [black, above=9, left=2] {$B_1$};
				\draw [black, line width=.55mm] (5,0) rectangle (4,3) node [black, above=8, right=4] {$B_2$};
				\draw [black, line width=.55mm] (2,-1.5) rectangle (3,-.5) node [black, below=37, left=4] {$B_3$};
				\draw [black,fill] (.6,1.95) circle [radius=0.079] node [black,left=2] {1};
				\draw [black,fill] (.6,1.05) circle [radius=0.079] node [black,left=2] {2};
				\draw [black,fill] (4.4,2.35) circle [radius=0.079] node [black,right=2] {3};
				\draw [black,fill] (4.4,1.45) circle [radius=0.079] node [black,right=2] {4};
				\draw [black,fill] (4.4,.55) circle [radius=0.079] node [black,right=2] {5};
				\draw [black,fill] (2.5,-.9) circle [radius=0.079] node [black,below=2] {6};
				\draw [blue, loosely dashed, thick] (.6,1.05)
				to (2.5,-.9);
				\draw [blue, loosely dashed, thick] (4.4,1.45)
				to (4.4,.55);
				\draw [red, densely dashed, thick] (.6,1.05)
				to (4.4,1.45);
				\draw [red, densely dashed, thick] (2.5,-.9)
				to (4.4,.55);
			\end{tikzpicture}
			\caption{A quadratic Markov move for the additive SBM with blocks $\mathbb{B}$.}
		\end{subfigure}
		\ \ \ \ \ 
		\begin{subfigure}{.3\textwidth}
			\begin{tikzpicture}
				\draw [black, line width=.55mm] (0,.5) rectangle (1,2.5) node [black, above=9, left=2] {$B_1$};
				\draw [black, line width=.55mm] (5,0) rectangle (4,3) node [black, above=8, right=4] {$B_2$};
				\draw [black, line width=.55mm] (2,-1.5) rectangle (3,-.5) node [black, below=37, left=4] {$B_3$};
				\draw [black,fill] (.6,1.95) circle [radius=0.079] node [black,left=2] {1};
				\draw [black,fill] (.6,1.05) circle [radius=0.079] node [black,left=2] {2};
				\draw [black,fill] (4.4,2.35) circle [radius=0.079] node [black,right=2] {3};
				\draw [black,fill] (4.4,1.45) circle [radius=0.079] node [black,right=2] {4};
				\draw [black,fill] (4.4,.55) circle [radius=0.079] node [black,right=2] {5};
				\draw [black,fill] (2.5,-.9) circle [radius=0.079] node [black,below=2] {6};
				\draw [black] (.6,1.05)
				to (4.4,.55);
				\draw [black] (4.4,2.35)
				to (2.5,-.9);
				\draw [black] (.6,1.95)
				to (4.4,.55);
				\draw [black] (.6,1.95)
				to (4.4,2.35);
				\draw [black] (.6,1.05)
				to (2.6,-.9);
				\draw [black, line width=.0003mm] (4.4,2.35)
				to (4.4,1.45);
				\draw [black] (4.4,1.45)
				to (4.4,.55);
			\end{tikzpicture}
			\caption{A graph $h$ with $T_{\mathrm{Add}}(h)=T_{\mathrm{Add}}(g)$.\ \ \ \ \ \ \ \ \ \ \ \ \ \ \ \ \ \ \ \ \ \ \ \ \ \ \ \ \ \ \ \ \ \ \ \ \ \ \ \ \ \ \ \ \ \ \ \ \ \ \ \ \ \ \ \ }
			\label{fig:betaSBMg1}
		\end{subfigure}
		\caption{Two graphs $g$ and $h$ in the same fiber of the additive SBM with block structure $\mathbb{B}=\{B_1,B_2,B_3\}$ and a quadratic Markov move corresponding to the binomial $x_{26}x_{45}-x_{24}x_{56}\in\ker(\varphi_{\mathrm{Add}})$ that moves from $g$ to $h$.  The blue, loosely dashed lines indicate edge insertion and the red, densely dashed lines indicate edge deletion.}
		\label{fig: markov basis for additive SBM}
	\end{center}
\end{figure}

\subsection{The $\beta$-SBM}
\label{sec:alg:BetaSBM}
The $\beta$-SBM considers a partition of the node set $[n]$ into nonempty blocks $\B := \{B_1,\ldots,B_k\}$ and assigns each edge $g_{ij}$  a probability $p_{ij}$ that depends on both the nodes $i$ and $j$ as well as the blocks $B_{z(i)}$ and $B_{z(j)}$ that contain them.  
From Definition~\ref{def:betaSBM} 
we can quickly see that the sufficient statistics for this model consist of the degree of each node $i\in[n]$ together with the number of edges between each pair of blocks $B_i$ and $B_j$ for $1\leq i \leq j\leq k$.  
In other words, if we let $T_{\mbox{\tiny ER}}: \mathbb{G}\longrightarrow \R^{k+1\choose 2}$ and $T:\mathbb{G} \longrightarrow \R^n$ denote the sufficient statistics for the ER-SBM and $\beta$-model, respectively, then the sufficient statistics $T_\beta(g)$ for the $\beta$-SBM are given by the product map
\begin{equation*}
	\begin{split}
		T_{\mbox{\tiny ER}}\times T : \mathbb{G}\longrightarrow \R^{{k+1\choose 2}+n};\\
		T_{\mbox{\tiny ER}}\times T : g\longmapsto (T_{\mbox{\tiny ER}}(g),T(g)).\\
	\end{split}
\end{equation*}
From this description of the sufficient statistics, we can recover an $H$-representation for the model polytope $P_{\M_\beta}$ of the $\beta$-SBM.  
Since the sufficient statistics for the $\beta$-SBM is a product of the sufficient statistics for the ER-SBM and the $\beta$-model, it follows that the model polytope for $\beta$-SBM, $\mathcal{M}_\beta$, is a ``geometric'' product of the model polytopes of the ER-SBM and $\beta$-models.  
In the discrete geometry literature, this geometric product is called the \emph{free sum} of the two polytopes.  
A precise description of this operation and the proof of the  following result is given in Appendix C. 

\begin{prop}
	\label{prop: beta-SBM model polytope}
	Let $\B = \{B_1,\ldots,B_k\}$ be a partition of the node set $[n]$ into $k$ nonempty blocks.  
	Let $\M_\beta$, $\M_{\mbox{\tiny ER}}$, and $\M$ respectively denote the $\beta$-SBM, ER-SBM, and $\beta$-models with respect to $\B$ and $[n]$.  
	The model polytope $P_{\M_{\beta}}$ of the $\beta$-SBM is the free sum 
	$$
	P_{\M_{\beta}} = P_{\M_{\mbox{\tiny ER}}}\oplus P_{\M}.
	$$
\end{prop}

Combining Proposition~\ref{prop: beta-SBM model polytope} with standard techniques in discrete geometry, we can recover an $H$-representation of $P_{\mathcal{M}_\beta}$.  
However, unlike the ER-SBM and the additive SBM models, this $H$-representation is unfortunately exponentially-sized in the number of parameters.  

Finally, it remains to identify a Markov basis for the $\beta$-SBM $\M_{\beta}$.  
Markov moves for the $\beta$-SBM are naturally more complex, as they are formed by combining the moves of the $\beta$-model, which preserve node degrees, and the ER-SBM moves, which preserve the number of edges between and within blocks. 
While a complete theoretical description of a minimal Markov basis is an open problem specified in Appendix C, \cite{GPS21+}   implement a dynamic algorithm to generate applicable moves on the fibers of the $\beta$-SBM.  Notably, the results cited therein guarantee rapid mixing of the algebraic Markov chain for many fibers of this model.  
For all of our simulations in Sections~\ref{sec:PowerAndComparison} and \ref{sec:simulations}, standard diagnostics were used to test convergence of the chain; for example the {\tt CODA} package in {\tt R}, using trace plots, effective sample size, and Gelman-Rubin statistics. 
The construction of the moves relies on the so-called parameter hypergraph, defined in \cite[Section 2.2]{GPS16}. 

Theoretically, the desired Markov basis is encoded using a collection of binomials, analogous to the Markov bases described for ER-SBM and the additive SBM.  
The exact details of how to recover these binomials is given in the Appendix.  
Experimentally, this collection of binomials appears to consist of only cubics and quadratics.  
Similar to the moves described for the additive SBM model, the quadratic binomials encode Markov moves that exchange one pair of edges in a $4$-cycle between blocks for another.  
The cubics on the other hand, appear to correspond to exchanging a triple of edges spanning three blocks for another such triple. The following conjecture describes the set of moves for the $\beta$-SBM:
\begin{conj}
	A Markov basis for $\M_{\beta}$ consists of quadratic and cubic binomials, that is exchanges of edges 2 or 3 at a time. 
\end{conj}  
We end this section with a brief example of these binomials and an illustration of their corresponding moves.  

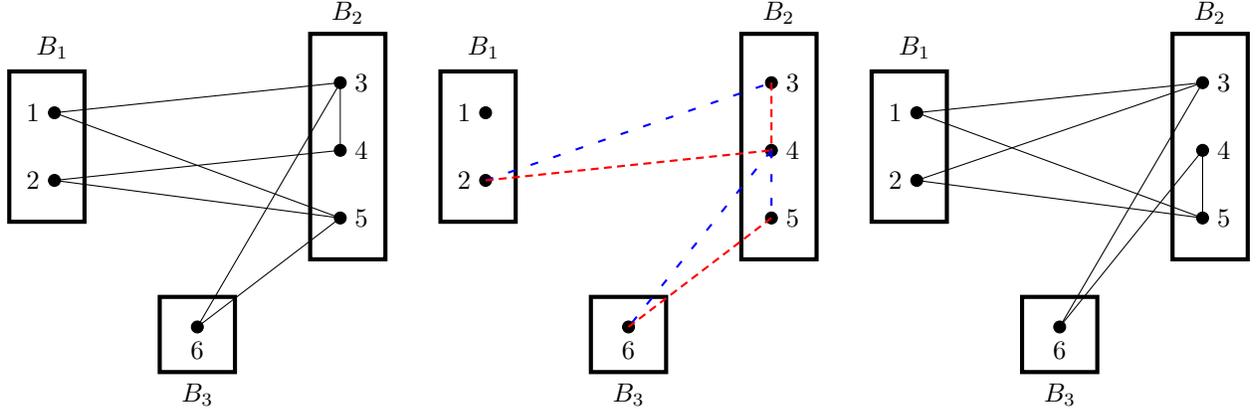
\begin{figure}[t!]
	\begin{center}
		\begin{subfigure}{.3\textwidth}
			\begin{tikzpicture}
				\draw [black, line width=.55mm] (0,.5) rectangle (1,2.5) node [black, above=9, left=2] {$B_1$};
				\draw [black, line width=.55mm] (5,0) rectangle (4,3) node [black, above=8, right=4] {$B_2$};
				\draw [black, line width=.55mm] (2,-1.5) rectangle (3,-.5) node [black, below=37, left=4] {$B_3$};
				\draw [black,fill] (.6,1.95) circle [radius=0.079] node [black,left=2] {1};
				\draw [black,fill] (.6,1.05) circle [radius=0.079] node [black,left=2] {2};
				\draw [black,fill] (4.4,2.35) circle [radius=0.079] node [black,right=2] {3};
				\draw [black,fill] (4.4,1.45) circle [radius=0.079] node [black,right=2] {4};
				\draw [black,fill] (4.4,.55) circle [radius=0.079] node [black,right=2] {5};
				\draw [black,fill] (2.5,-.9) circle [radius=0.079] node [black,below=2] {6};
				\draw [black] (.6,1.05)
				to (4.4,.55);
				\draw [black] (4.4,2.35)
				to (2.5,-.9);
				\draw [black] (.6,1.95)
				to (4.4,.55);
				\draw [black] (.6,1.95)
				to (4.4,2.35);
				\draw [black] (.6,1.05)
				to (4.4,1.45);
				\draw [black, line width=.0003mm] (4.4,2.35)
				to (4.4,1.45);
				\draw [black] (2.5,-.9)
				to (4.4,.55);
			\end{tikzpicture}
			\caption{A graph $g$ with blocks $\mathbb{B}$ and $T_{\beta}				(g)=(0,4,0,1,2,0,2,2,3,4,3,2)$.}
			\label{fig:betaSBMg1}
		\end{subfigure}
		\ \ \ \ \
		\begin{subfigure}{.3\textwidth}
			\begin{tikzpicture}
				\draw [black, line width=.55mm] (0,.5) rectangle (1,2.5) node [black, above=9, left=2] {$B_1$};
				\draw [black, line width=.55mm] (5,0) rectangle (4,3) node [black, above=8, right=4] {$B_2$};
				\draw [black, line width=.55mm] (2,-1.5) rectangle (3,-.5) node [black, below=37, left=4] {$B_3$};
				\draw [black,fill] (.6,1.95) circle [radius=0.079] node [black,left=2] {1};
				\draw [black,fill] (.6,1.05) circle [radius=0.079] node [black,left=2] {2};
				\draw [black,fill] (4.4,2.35) circle [radius=0.079] node [black,right=2] {3};
				\draw [black,fill] (4.4,1.45) circle [radius=0.079] node [black,right=2] {4};
				\draw [black,fill] (4.4,.55) circle [radius=0.079] node [black,right=2] {5};
				\draw [black,fill] (2.5,-.9) circle [radius=0.079] node [black,below=2] {6};
				\draw [blue, loosely dashed, thick] (.6,1.05)
				to (4.4,2.35);
				\draw [blue, loosely dashed, thick] (4.4,1.45)
				to (4.4,.55);
				\draw [blue, loosely dashed, thick] (4.4,1.45)
				to (2.5,-.9);
				\draw [red, densely dashed, thick] (.6,1.05)
				to (4.4,1.45);
				\draw [red, densely dashed, thick] (4.4,2.35)
				to (4.4,1.45);
				\draw [red, densely dashed, thick] (2.5,-.9)
				to (4.4,.55);
			\end{tikzpicture}
			\caption{A cubic Markov move for the $\beta$-SBM with blocks $\mathbb{B}$. \ \ \ \ \ \ \ \ \ \ \ \ \ \ \ \ \ \ \ \ \ \ \ \ \ \ \ \ \ \ \ \ \ \ \ \ \ \ \ \ \ \ \ \ \ }
		\end{subfigure}
		\ \ \ \ \
		\begin{subfigure}{.3\textwidth}
			\begin{tikzpicture}
				\draw [black, line width=.55mm] (0,.5) rectangle (1,2.5) node [black, above=9, left=2] {$B_1$};
				\draw [black, line width=.55mm] (5,0) rectangle (4,3) node [black, above=8, right=4] {$B_2$};
				\draw [black, line width=.55mm] (2,-1.5) rectangle (3,-.5) node [black, below=37, left=4] {$B_3$};
				\draw [black,fill] (.6,1.95) circle [radius=0.079] node [black,left=2] {1};
				\draw [black,fill] (.6,1.05) circle [radius=0.079] node [black,left=2] {2};
				\draw [black,fill] (4.4,2.35) circle [radius=0.079] node [black,right=2] {3};
				\draw [black,fill] (4.4,1.45) circle [radius=0.079] node [black,right=2] {4};
				\draw [black,fill] (4.4,.55) circle [radius=0.079] node [black,right=2] {5};
				\draw [black,fill] (2.5,-.9) circle [radius=0.079] node [black,below=2] {6};
				\draw [black] (.6,1.05)
				to (4.4,.55);
				\draw [black] (4.4,2.35)
				to (2.5,-.9);
				\draw [black] (.6,1.95)
				to (4.4,.55);
				\draw [black] (.6,1.95)
				to (4.4,2.35);
				\draw [black] (.6,1.05)
				to (4.4,2.35);
				\draw [black] (4.4,1.45)
				to (4.4,.55);
				\draw [black] (4.4,1.45)
				to (2.5,-.9);
				
			\end{tikzpicture}
			\caption{A graph $h$ with $T_{\beta}(h)=T_{\beta}(g)$. \ \ \ \ \ \ \ \ \ \ \ \ \ \ \ \ \ \ \ \ \ \ \ \ \ \ \ \ \ \ \ \ \ \ \ \ \ \ \ \ \ \ \ \ \ \ \ \ \ \ \ \ \ \ \ \ \ \ \ \ }
			\label{fig:betaSBMg1}
		\end{subfigure}
		\caption{Two graphs $g$ and $h$ in the same fiber of the $\beta$-SBM with block structure $\mathbb{B}=\{B_1,B_2,B_3\}$ and a cubic Markov move corresponding to the binomial $x_{23}x_{45}x_{46}-x_{24}x_{34}x_{56}\in\ker(\varphi_\beta)$ that moves from $g$ to $h$.  The blue, loosely dashed lines indicate edge insertion and the red, densely dashed lines indicate edge deletion.}
		\label{fig: markov move beta SBM}
	\end{center}
\end{figure}

\begin{example}
	\label{ex: beta-SBM example}
	Let $n := 6$ and partition $[n]$ into the nonempty blocks 
	$$
	\B := \{
	B_1:= \{1,2\},
	B_2 := \{3,4,5\},
	B_3 := \{6\}
	\}.
	$$
	With the help of {\tt Macaulay2} \citep{M2}, we compute the collection of binomials yielding our desired Markov basis by way of the algebraic methods described in the Appendix.  
	The result is the following collection of $18$ quadrics and $2$ cubics:
	\begin{equation*}
		\begin{split}
			\langle 
			&x_{35}x_{46}-x_{34}x_{56}, 
			x_{25}x_{46}-x_{24}x_{56}, 
			x_{15}x_{46}-x_{14}x_{56}, 
			x_{36}x_{45}-x_{34}x_{56}, \\
			&x_{25}x_{36}-x_{23}x_{56}, 
			x_{24}x_{36}-x_{23}x_{46}, 
			x_{15}x_{36}-x_{13}x_{56}, 
			x_{14}x_{36}-x_{13}x_{46}, \\
			&x_{24}x_{35}-x_{23}x_{45}, 
			x_{14}x_{35}-x_{13}x_{45}, 
			x_{25}x_{34}-x_{23}x_{45}, 
			x_{15}x_{34}-x_{13}x_{45}, \\
			&x_{16}x_{25}-x_{15}x_{26}, 
			x_{16}x_{24}-x_{14}x_{26}, 
			x_{15}x_{24}-x_{14}x_{25}, 
			x_{16}x_{23}-x_{13}x_{26}, \\
			&x_{15}x_{23}-x_{13}x_{25},
			x_{14}x_{23}-x_{13}x_{24}, \\
			&x_{23}x_{45}x_{46}-x_{24}x_{34}x_{56}, 
			x_{13}x_{45}x_{46}-x_{14}x_{34}x_{56}
			\rangle.
		\end{split}
	\end{equation*}
	The quadratic binomials above correspond to Markov moves like those captured in Theorem~\ref{thm: corollary to markov basis theorem for additive SBM} and depicted in Figure~\ref{fig: markov basis for additive SBM}.  
	The two cubic binomials correspond to new moves that are particular to the $\beta$-SBM.  
	Graphically, these ``cubic" moves correspond to an exchange of two collections of three edges spanning three blocks.  
	One of these cubic moves is illustrated in Figure~\ref{fig: markov move beta SBM}.  
\end{example}

\section{Small-sample power calculations and comparison with an existing  method} 
\label{sec:PowerAndComparison}

The following simulation experiments showcase the small-sample power of our tests and  compare them with the existing test of \cite{lei2016goodness}. We  also explore the effects that the 3 elements of the goodness-of-fit test have on its power: the choice of a test statistic that measures departures from model fit, 
the method for approximating the distribution of the test statistic, and the method of estimating the unknown block assignment $\mathcal Z$. 
For the test statistic, we have two choices: the block-corrected $\chi^2_{\mbox{Block-corrected}}(g,\blocks)$ distance defined in Equation~\eqref{eqn:DegreeCorrectedChiSquared}, \S~\ref{sec:GoFstatistic}, and the test statistic based on the spectral gap, which is the difference between the first and second eigenvalue of a residual matrix as defined in \cite{lei2016goodness}. There are three methods for approximating the distribution of the test statistic: a finite-sample approximation based on the Monte Carlo sampling of graphs from the fiber (called the \emph{fiber sampler}), an asymptotic approximation, and a bootstrap approximation; note that the last two approximations are applicable only for the spectral gap statistic as defined in \cite{lei2016goodness}. Finally, for the estimation of block assignments $\mathcal Z$, there are three choices: the spectral method, using the true block assignment, and the Bayes method of \cite{pati2015optimal}. 
The first two block assignments methods allow for a direct comparison with the state of the art.  

We simulate $B = 50$ graphs, each with $n = 27$ nodes from the $\beta$-SBM, with $7$ different parameter settings 
summarized in Table \ref{tab:paramsPower}. Parameter settings 1 to 4 correspond to a $\beta$-SBM that with high probability, generates networks with two well separated dense clusters with heterogeneous degrees. On the other hand, settings 5,6,7 with high probability, generate networks with sparse clusters. Because of the sparsity of clusters and small $n$, the degrees are not that heterogeneous. Due to this, graphs from settings 5, 6 and 7 can be approximated by an ER-SBM, even though they are generated from a $\beta$-SBM, making it a challenge for the goodness-of-fit test to detect departures from the ER-SBM.  

\begin{remark}
	For parameter settings 1 through 5, we use the log-linear parametrization to generate the graphs, i.e. $p_{uv} = \text{invlogit}(\alpha_{z_u z_v}+\beta_u + \beta_v) $. For settings 6 and 7, for comparison, we use the degree-corrected parametrization, i.e. $p_{uv} = e^{\beta_i}e^{\beta_j}e^{\alpha_{z_u z_v}}  $. We make this choice of parametrization since it is popular in the literature, although the log-linear parameterization is superior, see \cite{Xiaolin2015thesis} and Remark \ref{remark:dcparam}.
\end{remark}

\begin{table}[htbp]
	\centering
	\scalebox{0.85}{
		\begin{tabularx}{.80\textwidth}{l@{\hskip 15mm}l@{\hskip 8mm}l}
			\toprule
			&
			\multicolumn{2}{c}{\bf Parameters} \\ 
			\cmidrule{2-3}
			{\bf Parameter Setting} & 
			\multicolumn{1}{c}{$\alpha$} &
			\multicolumn{1}{c}{ $\beta$} \\
			\midrule
			\rule{0pt}{5ex}
			1 &
			$\alpha = \begin{pmatrix}
				0.6 & 0.1 \\
				0.1 & 0.3 
			\end{pmatrix}$ &
			$\beta_u \sim \mbox{Unif}(-\text{n,n})$ \\ 
			\rule{0pt}{5ex}
			2 &
			$\alpha = \begin{pmatrix}
				0.6 & 0.1 \\
				0.1 & 0.3 
			\end{pmatrix}$ &
			$\beta_u \sim \mbox{Unif}(-\text{10,10})$ \\
			\rule{0pt}{7ex}
			3 &
			$\alpha = \begin{pmatrix}
				-2 & -1 \\
				-1 & -0.01 
			\end{pmatrix}$ &
			$\beta_u \sim \mbox{Unif}(-\text{n,n})$ \\ 
			\rule{0pt}{5ex}
			4 &
			$\alpha = \begin{pmatrix}
				-2 & -1 \\
				-1 & -0.01 
			\end{pmatrix}$ &
			$\beta_u \sim \mbox{Unif}(-\text{10,10})$ \\ 
			\rule{0pt}{5ex}
			5 & 
			$\alpha = \begin{pmatrix}
				\log(0.6) & \log(0.1) \\
				\log(0.1) & \log(0.3) 
			\end{pmatrix}$ &
			$\beta_u \sim \log(\mbox{Unif}(0,1))$ \\	
			\rule{0pt}{5ex}
			$6^1$ & 
			$\alpha = \begin{pmatrix}
				\log(0.6) & \log(0.1) \\
				\log(0.1) & \log(0.3) 
			\end{pmatrix}$ &
			$\beta_u \sim \log(\mbox{Unif}(0,1))$\\ 
			
			\rule{0pt}{5ex}
			$7^1$ &
			$\alpha = \begin{pmatrix}
				\log(0.6) & \log(0.2) \\
				\log(0.2) & \log(0.6) 
			\end{pmatrix}$ &
			$\beta_u \sim \log(\mbox{Unif}(0,1))$ \\ 
			\bottomrule
			\multicolumn{3}{l}{ 
				$^1$Settings 6 and 7 are the degree corrected parametrization.} \\
			\multicolumn{3}{l}{Settings 1 through 5 are the log-linear parametrization.}
		\end{tabularx}
	}
	%
\vspace{5mm}
\caption{Model parameter values for simulated graphs from the $\beta$-SBM 
}
\label{tab:paramsPower}
\end{table}

We run 9 different goodness-of-fit tests and record the number of times each test rejects the null hypothesis that the network comes from an ER-SBM. As discussed below, various combinations of these nine tests show the effects of different test elements on the power.  The test choices are as follows (see Table \ref{tab:tests}): 
\begin{enumerate}
\item 
{Spectral asymptotic test with spectral gap.} This is a  test from \cite{lei2016goodness}. 
\item {Spectral bootstrap test with  spectral gap.} This is also a  test from \cite{lei2016goodness}.
\item {Spectral fiber test with $\chi^2$.} 
\item {Spectral fiber test with  spectral gap.} 
\item {True-blocks asymptotic test with spectral gap.} 
\item {True-blocks bootstrap  test with spectral gap.} 
\item {True-blocks fiber with $\chi^2$.} 
\end{enumerate} 
\begin{table}[htbp]
\centering
\begin{tabular}{c|c|c|c}
	\toprule
	& Test statistic & 		\multirow{2}{5cm}{Method to approximate the distribution of test statistic} & 	\multirow{2}{5cm}{Method to estimate $\mathcal Z$} \\
	&                &                                                               &                              \\
	\midrule
	\midrule
	Test 1 
	& Spectral gap &  Asymptotic & Spectral method \\
	Test 2 
	& Spectral gap &  Bootstrap-corrected & Spectral method \\
	Test 3 
	& $\chi^2_{BC}$ & Finite-sample from fiber &Spectral method \\
	Test 4 
	& Spectral gap & Finite-sample from fiber &Spectral method \\
	Test 5 
	& Spectral gap &Asymptotic &  True $\mathcal Z$\\
	Test 6 
	& Spectral gap &Bootstrap-corrected &True $\mathcal Z$\\
	Test 7 
	& $\chi^2_{BC}$ & Finite-sample from fiber &True $\mathcal Z$\\
	\bottomrule
\end{tabular}
\vspace{5mm}
\caption{The different goodness-of-fit tests we compared with varying test elements: the choice of a test statistic, 
	the method for approximating the distribution of the test statistic, and the method of estimating the unknown block assignment $\mathcal Z$.}
\label{tab:tests}
\end{table}

The setup of Tests 1, 2,  3, and 4 allow us to compare the small sample power of our tests based on the fiber sampler with the small sample power of \cite{lei2016goodness}, which are based on asymptotic and bootstrap approximations. For a fair comparison, we use the spectral method of estimating $\mathcal Z$ along with \cite{lei2016goodness}'s test statistic, as defined by the spectral gap, in combination with our fiber-sampler.  Results are summarized  in Table~\ref{tab:powercomparison}. 

\begin{table}[htbp]
\centering
\begin{tabular}{c|cc|cc}
	\toprule
	\multirow{2}{*}{Parameter Setting} &
	Test 1 & Test 2 & Test 3 & Test 4\\
	& {(Asymptotic)} & {(Bootstrap)} & {($\chi^2$ on Fiber)} & (Spectral gap on Fiber) \\
	\midrule
	\midrule
	1 & 1 		& 0.8 	& 1 	& 1 	 \\
	2 & 0.98 	& 0.74 	& 1 	& 0.98 	\\
	3 &	1 		& 0.78 	& 1 	&	1 		\\
	4 & 1 		& 0.82 	& 1 	& 0.98 	\\
	5 & 0.16	& 0.02	& 0.6	& 0.1 \\
	6 & 0.38 	& 0.02 	& 0.82 	& 0.18 	\\
	7 &	0.34 	& 0.04 	& 0.8 	& 0.3 	\\
	\bottomrule
\end{tabular}
\vspace{5mm}
\caption{Test power for $50$  runs of the the goodness-of-fit tests for the ER-SBM for $n=27$ nodes. Comparison  of the fiber-sampler method developed here with  an existing method from \cite{lei2016goodness}.}
\label{tab:powercomparison}
\end{table}
For small samples, tests that are based on the fiber sampler developed in \S~\ref{sec:SBMExactTest} (i.e., Tests 3 and 4) are more powerful than the test based on the asymptotic and bootstrap approximations from \cite{lei2016goodness} (i.e., Tests 1 and 2). This shows that for small graphs, the finite sample distribution of the test statistic, as estimated by the fiber sampler, offers a better approximation when compared to the bootstrap and the asymptotic methods of estimating the distribution. 

In comparing Tests 3 with Test 4, we note that both these tests use the fiber sampler to approximate the finite sample distribution. But Test 3, which uses with the $\chi^2$ test statistic, is more powerful than test 4, which uses the spectral gap test statistic, especially for parameter settings 5, 6, and 7. This provides evidence for the fact that our modified chi-square test statistic may be more powerful (in small samples) in detecting departures from the ER-SBM when compared to the spectral gap test statistic of \cite{lei2016goodness}. 
It may be worth noting that the spectral estimate of $\mathcal Z$ gave Test 3 (our fiber-sampler test in combination of the $\chi^2$ statistic) of those tests higher power \emph{even though}, as a matter of fact, that estimate was nowhere near the true block assignment! 

\smallskip 
Next, we are interested in evaluating the performance of the fiber sampler test when coupled with different methods for estimating the block assignment $\mathcal Z$ 	 that can be compared to \cite{lei2016goodness}: the spectral method and the true assignment $\mathcal Z$. 
To this end, we compare Tests 3, 4, 
5, 6, and 7; see the results in Table~\ref{tab:powerforfibersampler}. The choice of which method to estimate $\mathcal Z$ does not seem to effect the power in parameter settings 1 through 4. However, in settings 5 through 7, the spectral method to estimate $\mathcal Z$ seems to offer the highest small sample power, which curiously, is even better than the true $\mathcal Z$.

\begin{table}[htbp]
\centering
\begin{tabular}{c|cc|
		cc|ccc}
	\toprule
	\multirow{2}{*}{Parameter Setting} &
	Test 3 & Test 4 &  
	Test 5 & Test 6	 & Test 7\\
	& \multicolumn{2}{c}{Spectral $\mathcal Z$'s} &
	\multicolumn{3}{c}{True $\mathcal Z$'s} \\
	\cmidrule{2-6}
	& {$\chi^2$} & Eigenvalue & 
	Asymptotic & Boot & $\chi^2$ \\
	\midrule
	1 & 1 	& 1 	
	& 1 	& 1 	& 1 \\
	2 & 1 	& 0.98 	
	& 1 	& 1 	& 1 \\
	3 & 1 	&	1 	
	& 1 	& 1		&	1 	\\
	4 &1 	& 0.98 	
	& 1 	& 1 	& 1 \\
	5 & 0.6	& 0.1 
	& 0.3	& 0.02	& 0.28 \\
	6 & 0.82 	& 0.18 
	& 0.32 	& 0  	& 0.3 \\
	7 & 0.8 	& 0.3 
	& 0.38 	& 0.14 	& 0.54 \\
	\bottomrule
\end{tabular}
\vspace{5mm}
\caption{Power calculations over $50$  runs of the fiber-sampler goodness-of-fit test for the ER-SBM for $n=27$ nodes.}
\label{tab:powerforfibersampler}
\end{table}

\smallskip 
Tests 5, 6, and 7 allow us to additionally compare the power of the two test statistics (spectral gap and $\chi^2$), independently of the method of estimating the block assignment $\mathcal Z$. That is, if the true block assignment were known, we can measure how powerful a test statistic is to detect departures from the null without having to deal with the estimation of $\mathcal Z$; see Table~\ref{tab:powerforfibersampler}.
As can be seen from the table, the fiber-sampler tests perform better with the $\chi^2$ statistic, so we do not use the spectral gap statistics when the true $\mathcal Z$ is used. 
The table indicates that  the  power is (marginally) better with the $\chi^2$ statistic.

\paragraph{Power calculations for $\beta$-SBM.}
To practitioners who fit and use SBMs for data analysis, of most interest will likely be testing  goodness-of-fit of $\beta$-SBM when the network is actually generated by something else. To this end, we  report the power calculations for testing the null hypothesis that the network comes from the $\beta$-SBM with $k=k_1$ when in fact, its generated from a more general model with possibly a different $k=k_0$. We consider three different models to generate the network under the alternative hypothesis:
\begin{enumerate}
\item A dense $\beta$-SBM with $k_0 = 4$;
\item A sparse $\beta$-SBM with transitivity effects, with $k_0 = 4$; 
\item A mixed membership-SBM from  \cite{airoldi2008mixed}, with $k_0 = 4$. 
\end{enumerate} 
The parameter settings for each of these models is given below:
\begin{enumerate}
\item \textbf{Dense $\beta$-SBM:} $n$=200, $k_0 = 4$, $\beta_n = \mbox{uniform}(0,1)/n$, $\alpha =  \text{invlogit}(Q)$ where
$Q$ is a $k_0 \times k_0$ matrix of inter block and intra block edge probabilities, such that the diagonal elements of $Q= 0.8$ and off-diagonal elements of $Q = 0.2$.
\item \textbf{Sparse $\beta$-SBM with transitivity:}  $n$=200, $k_0 = 4$, $\beta_n = \mbox{uniform}(0,1)/n$, $\alpha =  \text{invlogit}(Q)$ where $Q$ is a $k_0 \times k_0$ matrix of inter block and intra block edge probabilities, such that the diagonal elements of $Q= 0.02$ and off-diagonal elements of $Q = 0.01$. We also add an additional transitivity parameter corresponding to the sufficient statistic number of triangles, with coefficient = $2$. Note that we use the ERGM package \citep{hunter2008ergm} to generate such networks using MCMC sampling.
\item \textbf{Mixed membership-SBM:} $n=200$, $k_0 = 4$ with the $k_0$ dimensional mixed membership vector for each node $i$, given by $\pi_i \sim$ Dirichlet distribution with parameters $\{\text{uniform}(0,1)\}_{1}^K$,  the block memberships $\blocks_{i\rightarrow j} \sim \text{ Multinomial} (\pi_i)$ and  $\blocks_{j\rightarrow i} \sim \text{ Multinomial} (\pi_j)$, and each dyad is sampled with probability $p_{ij} \sim \text{ Bernoulli} (\blocks_{i \rightarrow j}^T Q \blocks_{j \rightarrow i})$ where $Q$ is a $k_0 \times k_0$ matrix of intra block and inter block edge probabilities with  the diagonal elements of $Q = 0.8$ and off-diagonal elements of $Q = 0.2$.
\end{enumerate}

For each of these settings, we generate $B=50$ networks from the alternate model and test the null hypothesis that the network comes from a $\beta$-SBM with $k_1 = 2,3,4$. We report the proportion of times the goodness-of-fit test rejects the null hypothesis. To estimate the block assignments for the sparse $\beta$-SBM with transitivity, we use the regularized spectral method proposed by \cite{qin2013regularized} which is proven to give consistent estimates for degree corrected SBMs. We use the block-corrected (BC) chi-square statistic (equation \ref{eqn:DegreeCorrectedChiSquared}) as the goodness-of-fit statistic.
The results of the power calculations are summarized in Table \ref{tab:betaSBMpower}.

\begin{table}[htbp]
\centering
\begin{tabular}{l|c|l|}
	\toprule
	Alternative & 		Null: $\beta$-SBM &Power\\
	\midrule
	\midrule
	Dense $\beta$-SBM, $k_0=4$   & $k_1 = 4$ &  ${6.3 \%}^{*}$ \\
	Dense $\beta$-SBM, $k_0=4$   & $k_1 = 3$ &  $82 \%$ \\
	Dense $\beta$-SBM, $k_0=4$   & $k_1 = 2$ &   $98 \%$\\
	\midrule
	Sparse $\beta$-SBM with transitivity, $k_0=4$   & $k_1 = 4$ & $70\%$  \\
	Sparse $\beta$-SBM with transitivity, $k_0=4$   & $k_1 = 3$ &  $100 \%$ \\
	Sparse $\beta$-SBM with transitivity, $k_0=4$   & $k_1 = 2$ & $100 \%$  \\
	\midrule
	Mixed Membership SBM, $k_0=4$   & $k_1 = 4$ & $72\%$\\   
	Mixed Membership SBM, $k_0=4$   & $k_1 = 3$ & $88\%$\\
	Mixed Membership SBM, $k_0=4$   & $k_1 = 2$ & $58\%$ \\
	\bottomrule
\end{tabular}
\vspace{5mm}
\caption{Power calculations for testing if a network comes from $\beta$-SBM for various alternatives. The values marked with $^*$ are the Type 1 errors.}
\label{tab:betaSBMpower}
\end{table}

The results illustrate that the test shows good power for all the three settings. For the dense $\beta$-SBM, the test has good performance giving a Type 1 error of $6.3\%$ and a high power when testing using the incorrect number of blocks. For more general models such as the sparse $\beta$-SBM with transitivity, and the mixed membership SBM, the test also shows good performance. 

\section{Simulations of the proposed latent-block test}
\label{sec:simulations}

In this section, we illustrate the performance of the finite-sample goodness-of-fit test from in the latent setup, described on page~\pageref{latent block assignment}, on synthetic datasets generated from the three variants of the SBM. We test the null hypothesis that the  ER-SBM with $k=2$ blocks fits the synthetic networks and evaluate Type I and II errors. In all experiments, we assume that block memberships are latent, but the number of blocks is known. The experiments proceed with the following steps.

\paragraph{Data generation:}
We generate synthetic networks from each of the three variants of the SBM from Definitions~\ref{def:BlockModel}, \ref{def:additiveSBM}, and \ref{def:betaSBM}, assuming that the data come from a network with $k=2$ blocks.  
The block assignment $\mathcal{Z}$ is generated by sampling  ${1,...,k}$ uniformly and independently  for each node. We consider two different network sizes, $n=27$ and $n=90$. In addition, we study two probability regimes: in the \emph{dense} regime, the propensity of nodes to create edges is high; this is achieved by choosing higher values of model parameters. In the \emph{sparse} regime, nodes form edges with lower probability and the probability parameters are lower.

For each of the models used to generate a synthetic network, we specify parameters $p_{uv}$ and sample edges as independent Bernoulli  variables with parameter $p_{uv}$. In particular, for ER-SBM and additive SBM, we specify a block matrix of probabilities. For $\beta$-SBM, we specify block parameters $\alpha_{ij}$ for $1\leq i,j\leq k$ and node parameters $\beta_u$ for $1\leq u\leq n$. The specific choices are displayed in Table \ref{tab:params}.

\begin{table}[htbp]
	\centering
	\scalebox{0.85}{
		\begin{tabularx}{.80\textwidth}{l@{\hskip 15mm}l@{\hskip 8mm}l}
			\toprule
			&
			\multicolumn{2}{c}{\bf Parameters} \\ 
			\cmidrule{2-3}
			{\bf Model for synthetic data} & 
			\multicolumn{1}{c}{Dense} &
			\multicolumn{1}{c}{Sparse} \\
			\midrule
			\rule{0pt}{5ex}
			ER-SBM &
			$Q = \begin{pmatrix}
				0.6 & 0.1 \\
				0.1 & 0.6 
			\end{pmatrix}$ &
			$Q = \begin{pmatrix}
				0.20 & 0.01 \\
				0.01 & 0.20 
			\end{pmatrix}$ \\ 
			\rule{0pt}{5ex}
			additive SBM &
			$Q = \begin{pmatrix}
				0.77 & 0.67 \\
				0.67 & 0.55 
			\end{pmatrix}$ &
			$Q = \begin{pmatrix}
				0.02 & 0.12 \\
				0.12 & 0.50 
			\end{pmatrix}$ \\
			\rule{0pt}{7ex}
			$\beta$-SBM & 
			$\alpha = \begin{pmatrix}
				0.6 & 0.1 \\
				0.1 & 0.3 
			\end{pmatrix}$ &
			$\alpha = \begin{pmatrix}
				-2.00 & -0.01 \\
				-0.01 & -1.00 
			\end{pmatrix}$ \\	
			\rule{0pt}{3ex}
			&
			\multicolumn{2}{c}{$\beta_u \sim \mbox{Unif}(-\text{n,n}), u= 1, \ldots, n$} \\
			\bottomrule
		\end{tabularx}
	}
	\vspace{5mm}
	\caption{Model parameter values for generation of synthetic datasets 
	}
	\label{tab:params}
\end{table}

\paragraph{Block assignment estimation:}
A key ingredient in the finite sample tests with latent block assignments is the estimation of the block assignment distribution $\mathcal Z$ in Step~\ref{step:estimateBlocks} of Algorithm~\ref{algo:latentSBMtestGeneral}. We follow a Bayesian approach \citep{pati2015optimal,geng2016probabilistic} which involves fitting an ER-SBM by placing uniform priors on the edge probabilities and a Dirichlet-Multinomial prior on the block assignments (refer to (7) - (10) in \cite{pati2015optimal}).  Markov chain Monte Carlo is used to sample from the posterior distribution of the block assignments by  simultaneously generating block, parameter and assignment sampling distributions in a three-dimensional Gibbs sampler.  $2000$ block assignment are collected to adequately represent the posterior distribution.  
\begin{figure}[t!]
	\centering
	\includegraphics[scale=1]{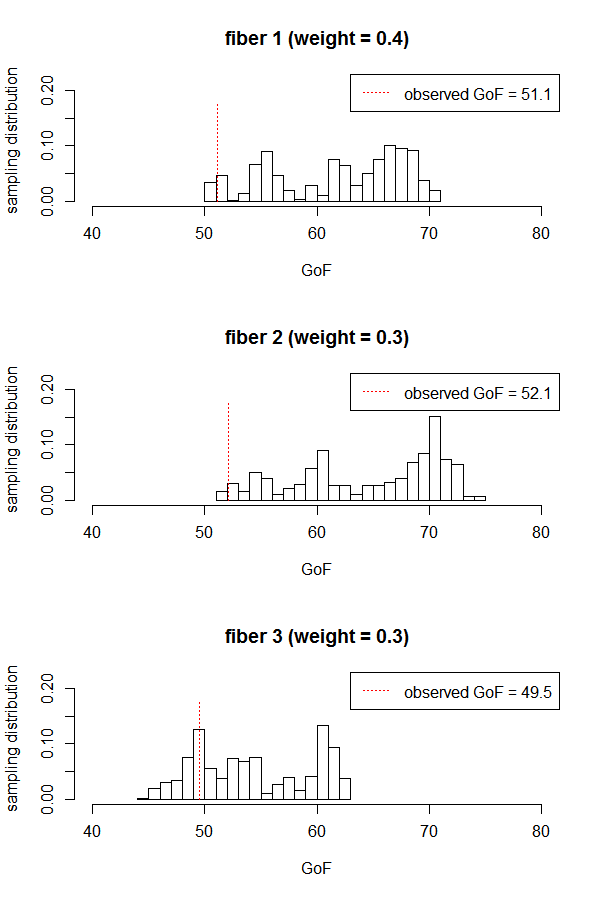} 
	\caption{Histograms of the GoF statistic for testing the fit of ER-SBM with $2$ blocks. Data generated from a $2$-block additive SBM. 
	}
	\label{fig:accept}
\end{figure}

\begin{figure}
	\centering
	\includegraphics[scale=.9]{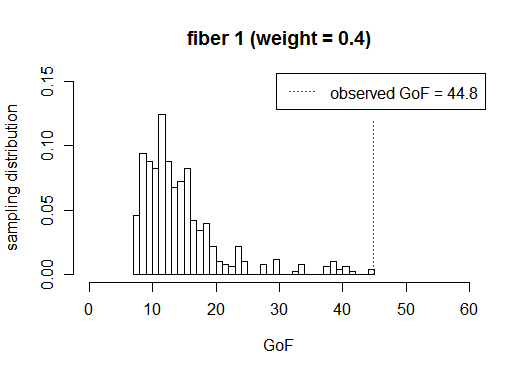}
	\quad
	\includegraphics[scale=.9]{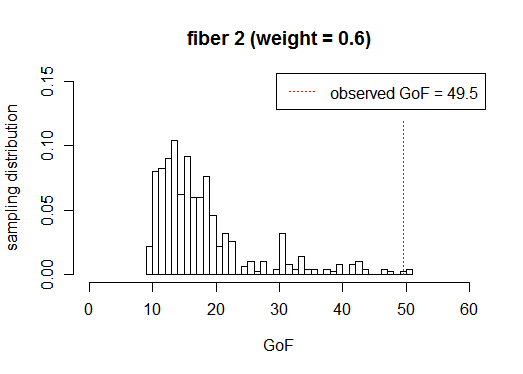}
	\caption{Histograms of the GoF statistic for testing the fit of a $2$-block ER-SBM. Data generated from a $2$-block $\beta$-SBM. 
	}
	\label{fig:reject}
\end{figure}

\paragraph{Fibers in the goodness-of-fit test:}
For each of the fibers in the posterior distribution, we generate the sampling distribution of the GoF statistic, as in Step~\ref{step:call algorithm for known blocks} in Algorithm~\ref{algo:latentSBMtestGeneral}.
Figures \ref{fig:accept} and \ref{fig:reject} demonstrate the inner workings of  these crucial ingredients of the goodness-of-fit test: they contain histograms of GoF statistics as generated on some of the fibers in a posterior distribution of $\mathcal Z$. (Of course, we do not plot $2000$ histograms; we select those with highest posterior probabilities.)  Each fiber is then used to compute a $p$-value (see Line~\ref{step:call algorithm for known blocks}, computing the $j$-the $p$-value). The final composite $p$-value is obtained by taking a weighted average of the $p$-values on all of the fibers, as in Step~\ref{step:avgpValueComputation} of Algorithm~\ref{algo:latentSBMtestGeneral},  using posterior probabilities as weights. This is then reported as the $p$-value of the goodness-of-fit test (see related discussion at the closing paragraphs of the Introduction in Section~\ref{sec:introduction}).

Figure \ref{fig:accept} illustrates three such fibers of the ER-SBM on $k=2$ blocks.  In this experiment, the synthetic data was generated using an additive SBM on  $k=2$ blocks.   The number of nodes is $n=90$ and the data is generated using a sparse parameter regime. The three  histograms of the sampling distributions of the GoF statistic shown correspond to the three fibers with highest posterior distribution; the fiber weights are re-scaled after  discarding the fibers whose posterior probability was only $1/2000$. The red dotted lines represent the observed GoF statistics.   The weighted $p$-value for this case is $0.87$ and the test does not reject the null.

Similarly, we tested goodness-of-fit of the $2$-block ER-SBM on synthetic data generated from the $2$-block $\beta$-SBM.  Figure \ref{fig:reject} shows sampling distributions of the GoF  statistic for two of the ER-SBM fibers with positive (non-trivial) posterior probabilities. 
The number of nodes in this experiment is $n=27$ and the networks are generated using a dense parameter regime. 
The test's $p$-value for this distribution is $0.005$ and therefore the null is rejected, i.e., the ER-SBM does not fit the synthetic data, as expected. 

\paragraph{The experiment and Type I/II error rates:}
Synthetic networks are generated 50 times from the null model and the block assignment estimation process is repeated for each network.  For each of these, the fibers are analyzed as described above. 

The proportion of rejections of the null hypothesis (i.e., conclusion of poor fit of the ER-SBM) at a nominal level of $0.05$ is reported  in Table \ref{tab:results}. 
As expected, when the synthetic networks are generated from the ER-SBM, the test rejects only $4 \%$ and $2 \%$ of the times for the dense and sparse case, respectively. 
When the networks are generated from the additive SBM, for the small networks with $n=27$, in both the dense and the sparse regime, the reported rejection rate is less than the nominal level of $0.05$. This is the expected result since the additive SBM is a sub-model of the ER-SBM and we expect not to reject the null. 
However, for the bigger networks with $n=90$, the test rejects in $8 \%$ and $14 \%$ of the samples, for the dense and sparse regimes, respectively.  
Finally, the test demonstrates good power, as seen by the results in the last rows of the Table: when the networks are generated from the $\beta$-SBM, for both $n=27$ and $n=90$ and in both the dense and the sparse regime, the test rejects the ER-SBM for 100\% of the samples.

\begin{table}[htbp]
	\centering
		
		\scalebox{0.9}{
			\begin{tabularx}{.78\textwidth}{l@{\hskip 15mm}l@{\hskip 15mm}l@{\hskip 15mm}l} \toprule
				
				&
				&
				\multicolumn{2}{c}{\bf Density} \\ 
				\cmidrule{3-4}
				{\bf Model for synthetic data} & 
				\multicolumn{1}{l}{Size} &
				\multicolumn{1}{l}{Dense} &
				\multicolumn{1}{l}{Sparse} \\
				\midrule
				\rule{0pt}{3ex}
				ER-SBM &
				$n=27$ &
				0.04 &
				0.02 \\
				&
				$n=90$ &
				0.04 &
				0 \\
				\rule{0pt}{3ex}
				additive SBM &
				$n=27$ &
				0 &
				0.04 \\
				&
				$n=90$ &
				0.08 & 
				0.14 \\
				\rule{0pt}{3ex}
				$\beta$-SBM &
				$n=27$ &
				1 &
				1 \\
				&
				$n=90$ &
				1 &
				1 \\
				\bottomrule
			\end{tabularx}
		}
		%
	\vspace{5mm}
	\caption{Rejection rates, at nominal level of $0.05$, over $50$  runs of the the goodness-of-fit test for the ER-SBM.
	}
	\label{tab:results}
\end{table}

\subsection{Application on real datasets} 
\label{sec:application}

We consider the application of our test to two real datasets: The \emph{Karate} dataset and the \emph{Human Connectome data}. 

\begin{figure}[htp!]
	\centering
	\includegraphics[width=7cm,height=6cm,keepaspectratio]{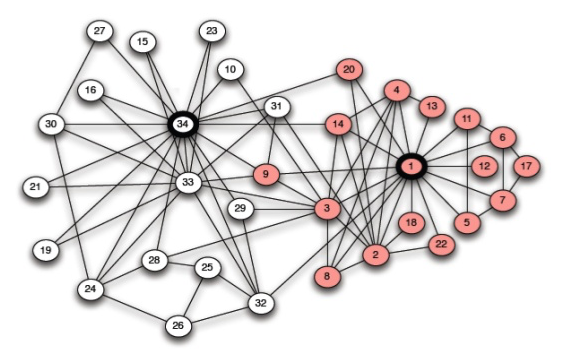}
	\caption{ Karate club data: reference  clustering}
	\label{fig:karatedata}
\end{figure}

\paragraph{Karate Club Network:} Zachary's Karate club data is a classic, well-studied social network of friendships between $34$ members of a Karate club at a US university in the 1970s \cite{zachary1977information}. The block structure of the Karate dataset is well understood: the network has 2 natural blocks (refer to Figure \ref{fig:karatedata}) that formed due to a social conflict that occurred during the observation; because of this reason it been used as a benchmark for community detection and fitting stochastic block models. 

\begin{figure}[t!]
	\centering
	\begin{subfigure}[b]{0.5\textwidth}
		\centering
		\includegraphics[height=4.2in,keepaspectratio]{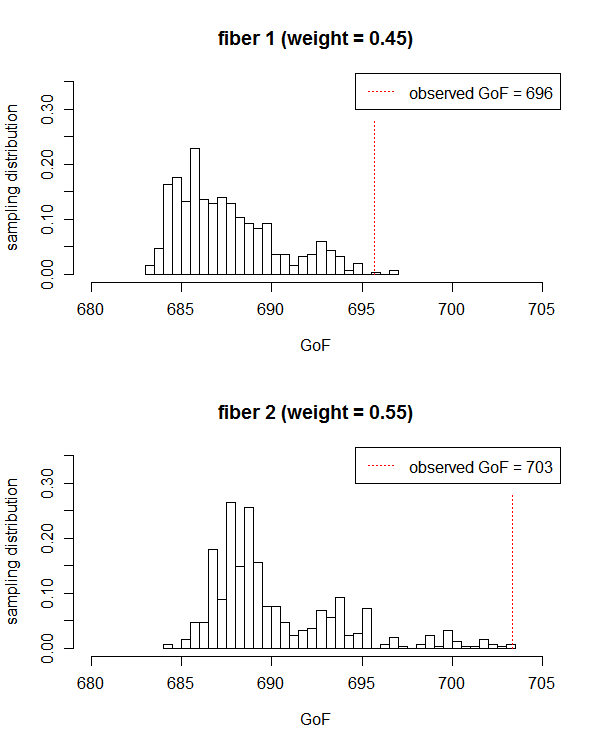}
		\caption{$k=2$, test rejects the null with $p$-value 0.01.}
	\end{subfigure}%
	~ 
	\begin{subfigure}[b]{0.5\textwidth}
		\centering
		\includegraphics[height=4.2in,keepaspectratio]{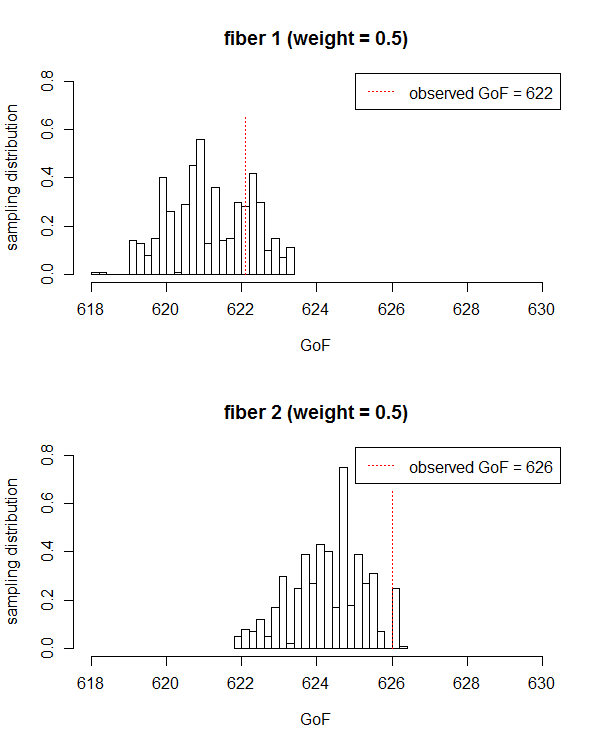}
		\caption{$k=4$, test favors the null with $p$-value 0.17.}
	\end{subfigure}
	\caption{Histograms of the GoF statistic for testing the fit of the Karate Club Network. Tested against the null hypothesis of ER-SBM with a) $2$ and b) $4$ blocks. 
	}
	\label{fig:karate}
\end{figure}

Some authors have also fit a SBM with $k=4$ blocks for the Karate dataset, see for e.g \cite{bickel2009nonparametric}.  We test the Karate dataset 
for the ER-SBM with $k=2$ and $k=4$ blocks. For $k=2$ blocks, we obtain a $p$-value of $0.01$ whereas for $k=4$ blocks, we obtain a $p$-value of $0.17$. Thus the tests rejects the $2$-block SBM but fails to reject the $4$-block SBM.  Histograms of the test statistic for the dominant fibers are demonstrated in Figure \ref{fig:karate}.   Each histogram in Figure \ref{fig:karate}  represents a fiber from a posterior distribution of block assignments. We display 2 fibers with highest posterior probabilities (re-scaled after  discarding those with probability was only $1/2000$). The red dotted lines represent the observed GoF statistics.

\cite{bickel2009nonparametric} noted that if no constraints are imposed on the edge probabilities, the Newman-Girvan modularity \cite{girvan2002community} and the likelihood-based modularity \cite{bickel2009nonparametric} provide conflicting estimates of the block assignments.  While Newman-Girvan modularity delivers a clustering result which seems to mostly agree with the split in Figure \ref{fig:karatedata}, the blocks obtained by the   likelihood based modularity  are quite different \cite{bickel2009nonparametric}.  It shows one  block consisting of five individuals with central importance that connect with many other nodes while the other block consists of the remaining individuals.  
However, under the constraint that within-block density is no less than the density of relationship to all other blocks, they both seem to match with the split in Figure \ref{fig:karatedata}.  Since the model fitting in \cite{pati2015optimal} does not impose any constraints in the edge probabilities, we expect our likelihood-based method to have a  poor fit for a 2-block SBM.    \cite{bickel2009nonparametric}  argued that 4-block SBM might be a better fit for the Karate network. This is indeed shown by the $p$-values obtained from the goodness-of-fit method. 

\begin{figure}
	\centering
	\begin{subfigure}{.5\textwidth}
		\centering
		\includegraphics[width=.65\linewidth]{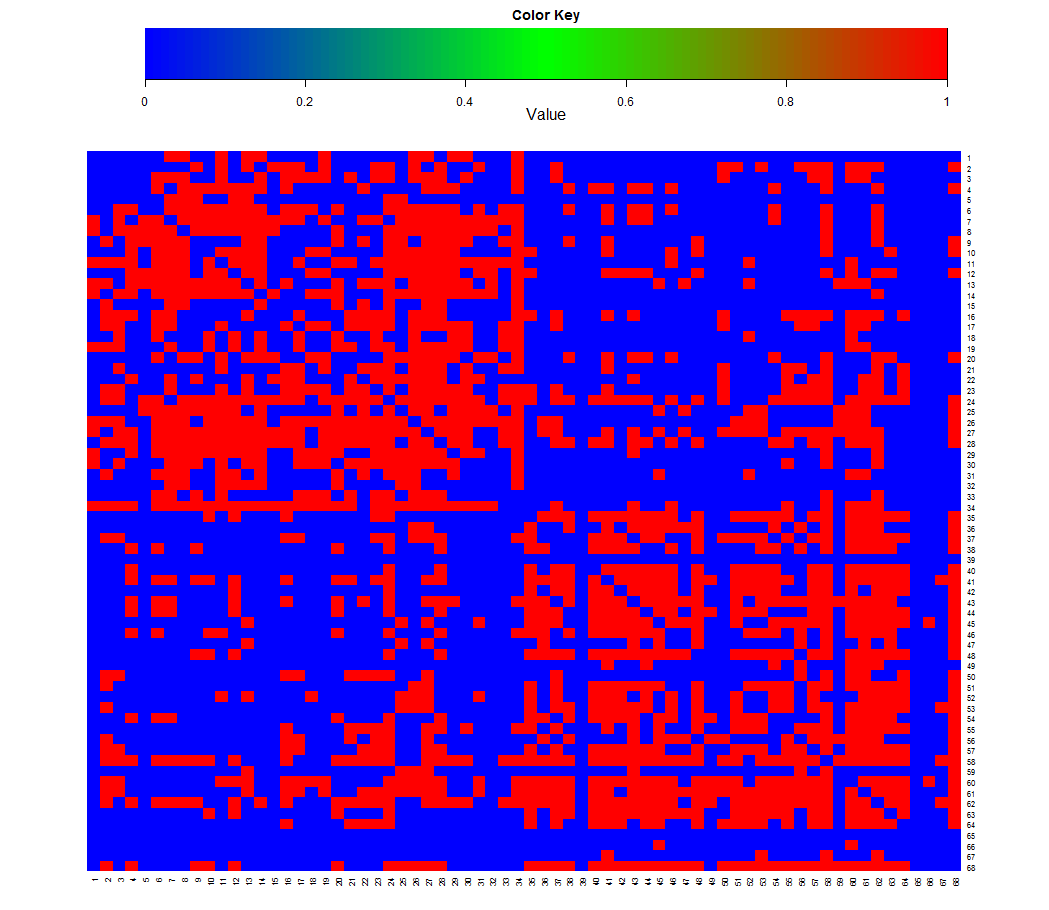}
		\caption{Subject 1}
		\label{fig:sub1}
	\end{subfigure}%
	\begin{subfigure}{.5\textwidth}
		\centering
		\includegraphics[width=.65\linewidth]{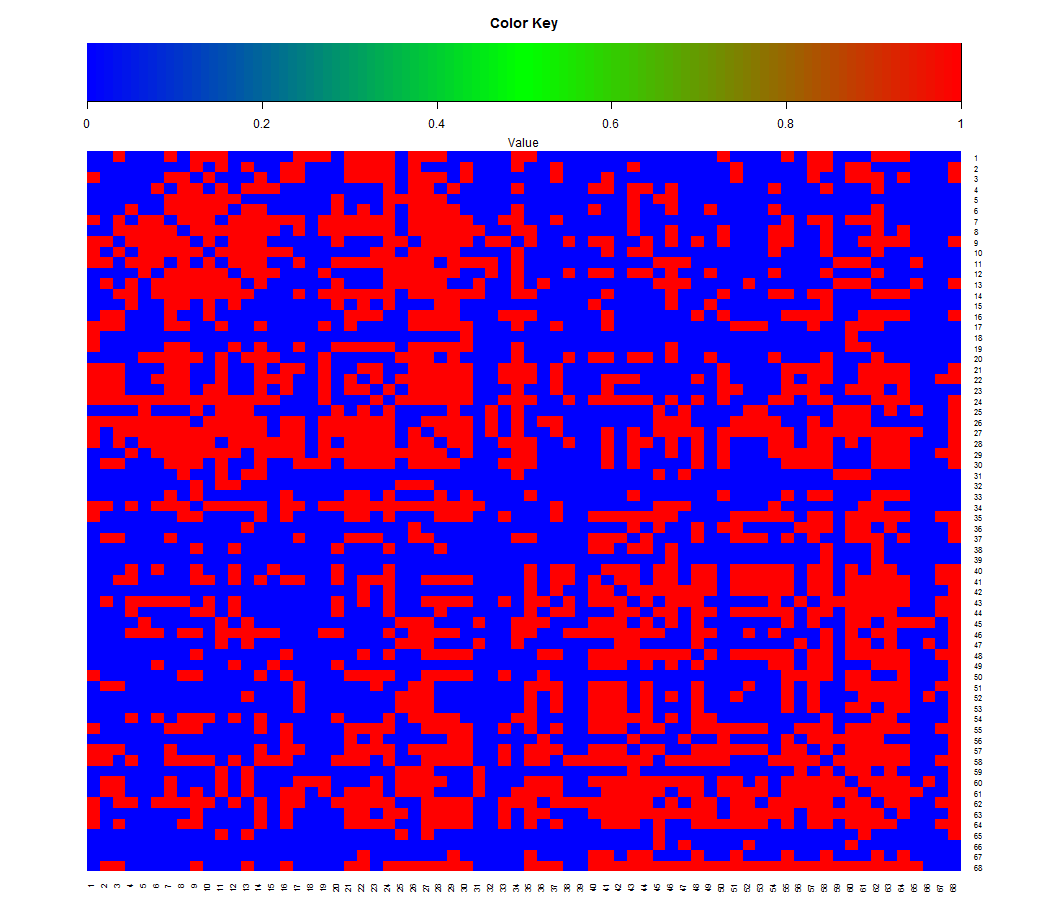}
		\caption{Subject 8}
		\label{fig:sub2}
	\end{subfigure}
	\caption{Heatmap of the adjacency matrices; blue pixels indicate 0 and red pixels indicate 1}
	\label{fig:hcp}
\end{figure}

\paragraph{Human Connectome Network:} We analyze the connections in the human brain, using the data provided by the Human Connectome Project, \url{http://www.humanconnectomeproject.org/}. \footnote{Data collection and sharing for this project was provided by the MGH-USC Human Connectome Project (HCP; Principal Investigators: Bruce Rosen, M.D., Ph.D., Arthur W. Toga, Ph.D., Van J. Weeden, MD). HCP funding was provided by the National Institute of Dental and Craniofacial Research (NIDCR), the National Institute of Mental Health (NIMH), and the National Institute of Neurological Disorders and Stroke (NINDS). HCP data are disseminated by the Laboratory of Neuro Imaging at the University of California, Los Angeles.
}
The data were pre-processed following the steps described in \cite{zhang2019tensor} to obtain a binary network representation. For each of $857$ subjects, the connectome is represented as a network on $68$ nodes, corresponding to two brain regions, the left and the right hemisphere with $34$ nodes in each hemisphere. If two regions demonstrate a high correlation between the fMRI signals (see \cite{Glasser2013105} for a detailed pipeline), then the corresponding nodes are linked by an edge. Thus, the choice of 2 blocks is natural in these data because of the left and the right hemisphere of the brain. For more detailed exposition, we consider two randomly chosen subjects: Subject 1 and Subject 8 and test for the null hypothesis that the data comes from an ER-SBM with $k=2$ blocks.   The heatmap for the adjacency matrices for Subjects 1 and 8 are shown in Figure \ref{fig:hcp}. 

\begin{figure}[t!]
	\centering
	\includegraphics[scale=1]{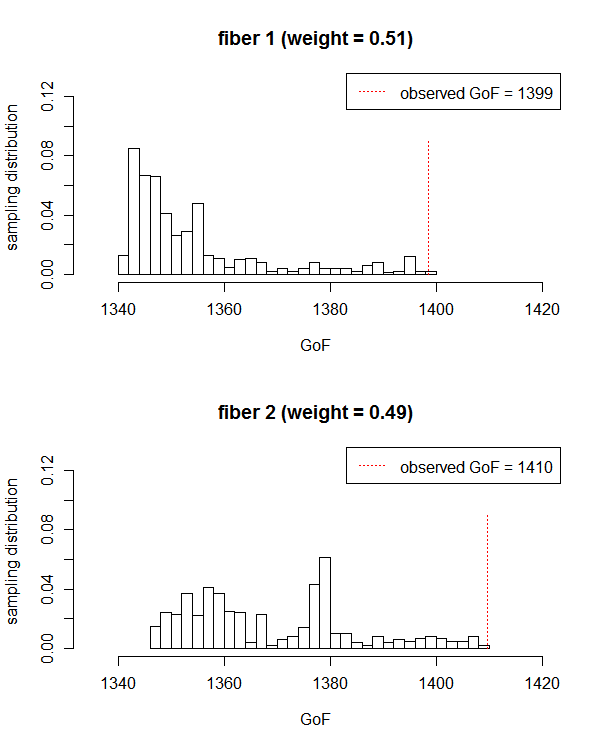}
	\caption{Histograms of the GoF statistic for testing the fit of the Human Connectome Network for subject 1. Tested against the null hypothesis of ER-SBM with $2$ blocks. 
	}
	\label{fig:brain}
\end{figure}

We obtained a $p$-value of $0.00025$ and $0.00059$ for Subject $1$ and $8$ respectively, thus rejecting the null hypothesis that the data comes from an ER-SBM with $k=2$ blocks.  We also implemented our goodness-of-fit test with $k=3,\ldots, 8$ values.  However, all the tests reject the null hypothesis. This suggests  that the connectomics data cannot  be adequately described by this SBM.  This is not surprising,  since the SBM assumes that the degree distribution of the different nodes to be the same. 
However, the human brain is highly complex with varying degree distributions for the regions.  To that end, we implemented a Bayesian version of the $\beta$-SBM as in \cite{NewmanReinert2016EstimatingNumBlocks} with unknown number of blocks with the fiber walk implementation from \cite{GPS16}.    Simulation results for the latent-block $\beta$-SBM show that the  $p$-values for Subjects $1$ and $8$ are $0.98$ and $0.79$, respectively, indicating good fit of the model.
In Subject $1$, there was one significant block assignment in the estimate with  $k=5$.  In Subject $8$,  the posterior block distribution $\blocks$ consisted of close to $2000$ fibers, those with weights $>0.01$ were included to produce the  weighted $p$-value above.  Samples from the two significant fibers are shown in Figure~\ref{fig:brain}. Each histogram in Figure \ref{fig:brain} represents a fiber from a posterior distribution of block assignments. We display 2 fibers with highest posterior probabilities (re-scaled after  discarding those with probability was only $1/2000$). The red dotted lines represent the observed GoF statistics. The test does rejects the null with $p$-value 0.

\begin{figure}[t!]
	\centering
	\includegraphics[scale=0.4]{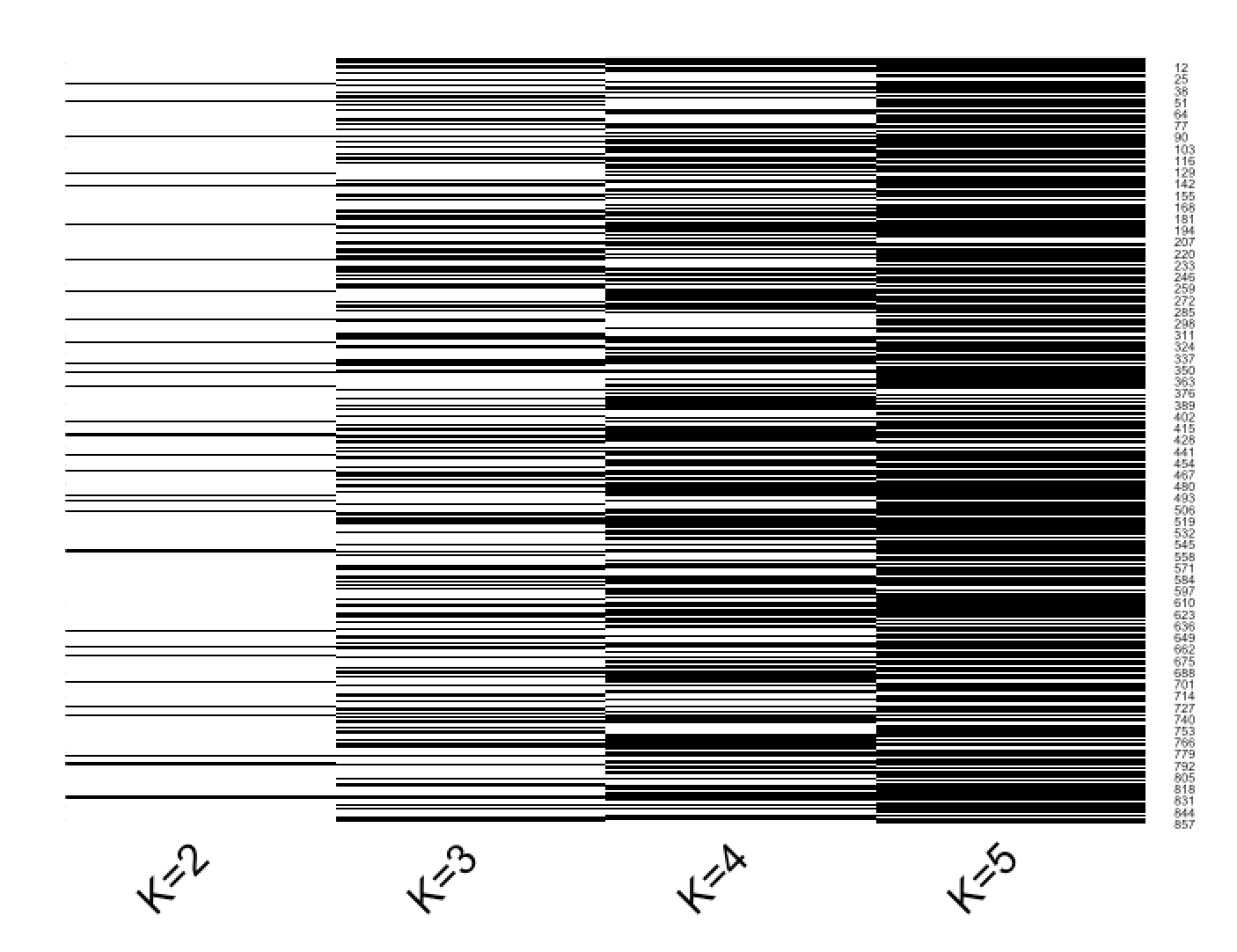}
	\caption{Binary heatmap of indicator of p-value $> 0.05$ when  $\beta$-SBM is fit with $K=2,3,4$ and $5$ (column) for all 857 individuals (row).}
	\label{fig:brainheat}
\end{figure}

As a comprehensive summary, Figure~\ref{fig:brainheat} displays a binary heat map (with 0 (white) indicating a significant p-value and 1(black) indicating a non-significant p-value) of all 857 individuals in the connectomics data when a $\beta$-SBM is fit with $K=2,3,4$ and $5$.  A Benjamini-Hochberg procedure is used on the 857 p-values to control the false discovery rate. Adjusting the rate helps to control for the fact that sometimes small p-values (less than 5\%) happen by chance, which could lead  to incorrect rejection of the true null hypotheses.  As $K$ increases, the number of individuals for which $\beta$-SBM cannot be rejected increases. More specifically, the percentages for non-significant p-values are $0.07, 0.38, 0.56$ and $0.76$ respectively.   

Interpreting the simulation results on the brain connectome network data comes with a caveat: the actual data sets has counts of how many neuron fibers connect between a pair of brain regions. In our simulation, we have dichotomized this connectivity network to a 0/1 network, summarizing the absence/presence of connections. While this is standard practice in network analysis, it is interesting to note that fitting a \emph{multi-graph} $\beta$-SBM version might capture more subtleties in the data. From the point of view of algebraic statistics, this is an easy extension - in fact the Markov chain algorithm will move faster, for all it takes is removing the 0/1 edge multiplicity requirement when attempting to use a Markov move. Theoretically, it means modeling the full contingency table rather than the restricted 0/1 contingency table. In the interest of space, we do not carry out those additional simulations.

\section{Discussion}
\label{sec:discussion}
Stochastic blockmodels are central to the network modeling literature, which,  starting in the early 1980s,  has proliferated in the direction of model development and simulation. 
While formal general methods for testing model fit have  remained elusive, recent literature  has begun to address the implementation of goodness-of-fit algorithms for specific models or families of models. 
For example, \cite{lei2016goodness}  provides an asymptotic test for the basic variant of the stochastic blockmodel, which we call the ER-SBM, along with a non-asymptotic version. \cite{BanerjeeMa2017OptimalTests} construct a hypothesis test for the stochastic blockmodel vs. the Erd\"os-R\`enyi model, with optimality guarantees under certain degree growth conditions. As discussed in Section~\ref{sec:models}, when the block assignment is known, these are examples of exponential family random graph models (ERGMs), for which general quantitative methods for goodness-of-fit testing are still lacking. 
This is an issue known in the statistics literature, as summarized in \cite[\S 2.3.4]{KolaczykBook2017}. 
For stochastic blockmodels in particular, the state-of-the-art methods on goodness-of-fit tests rely on asymptotic justifications with assumption that may or may not hold in practice. Such assumptions make it difficult  to modify the asymptotic methodology to handle SBM model variants, most notably the degree-corrected SBM. 

In this paper,  we develop the first non-asymptotic goodness-of-fit tests for both sparse and dense SBMs, with either known or unknown block assignments, including the degree-corrected SBM. 
For the non-latent case, we take the frequentist approach to construct an exact conditional test  that applies for all SBM variants. It is a broadly applicable test because it relies on the geometry of ERGMs, and discrete exponential families in general. The workhorse behind the test is a tool well-known in algebraic statistics, a Markov basis, used for sampling from the conditional distribution given the observed value of the sufficient statistics. 
This algebraic statistics method has never been applied to latent-variable models before. To extend the method to latent variables, we take the Bayesian approach, define a conditional $p$-value, and derive two interpretations of it in Section~\ref{sec:testTheoryBayes}. The core idea behind the implementation of this goodness-of-fit test is that we use the known-blocks test to implement the latent-block test, combined with  any given method for estimating the block assignments. 
Each of these  conditional tests can be viewed as a generalization of Fisher's exact tests for contingency tables to network models, and latent variable models more generally. In summary, the methodology developed in this manuscript  extends to any mixture of log-linear models on discrete data, where the estimation of the mixture parameter parallels block estimation, while each log-linear model comes equipped with a finite and, in principle, computable Markov basis, so that  algebraic statistics theory applies in the Bayesian context we propose. 

Simulations in Section~\ref{sec:simulations} show that the tests control the Type 1 error at the advertised nominal level and possess good power. The most general model to which we applied the test is the degree-corrected SBM from \cite{karrer2011stochastic}, the exponential-family version of which is called $\beta$-SBM. While we are able to computationally provide a full Markov basis for each example we tested, the problem of deriving formally a structural result of the Markov bases of the $\beta$-SBM model is open. 

On the applied side, among the remaining open problems, we single out the need for  additional research on the human connectome networks.  Specifically,  it is a common belief among neuroscientists that multiple subsets of the regions of the brain work in unison to perform various tasks. We thus  conjecture that a degree-corrected  mixed-membership model may be a better model for the connectivity of a human  brain. 
Significant work remains to develop both a model testing and selection procedures that include models with overlapping blocks with heterogeneous degree distribution.

\paragraph{Funding}All authors were members of the  Network Models Working Group in Algebraic Statistics 2016: American Mathematical Society's Mathematical Research Community program supported by the  National Science Foundation grant DMS 1321794. Sonja Petrovi\'c and Dane Wilburne were partially supported  by AFOSR FA9550-14-1-0141; Sonja Petrovi\'c was additionally supported by DOE award \#1010629 and the Simons Foundation Collaboration Grant for Mathematicians \#854770; Debdeep Pati was partially supported by Office of Naval Research (ONR BAA 14-0001) and the National Science Foundation (NSF DMS 1613156);  Liam Solus was partially supported by an NSF Mathematical Sciences Postdoctoral Research Fellowship (DMS - 1606407); Vishesh Karwa was partially supported by NSF DMS 1947919. The authors declare no conflict of interests.

\paragraph{Acknowledgments}The authors are grateful to the American Mathematical Society's Mathematical Research Community (MRC) program supported by the National Science Foundation  under Grant Number DMS 1321794. The authors are also grateful to the Associate Editor and two anonymous referees whose constructive feedback greatly improved the paper.

\paragraph{Data availability} The data sets used for this study are publicly available: the karate dataset is available from R package ERGM, \citep{hunter2008ergm} and the connectome data is available from the Human Connectome Project. (\url{http://www.humanconnectomeproject.org/data/}) The connectome data were pre-processed as described in \cite{zhang2019tensor}.

\bibliographystyle{abbrvnat}

\bibliography{JRSSb_networks}

\begin{thebibliography}{64}
\providecommand{\natexlab}[1]{#1}
\providecommand{\url}[1]{\texttt{#1}}
\expandafter\ifx\csname urlstyle\endcsname\relax
  \providecommand{\doi}[1]{doi: #1}\else
  \providecommand{\doi}{doi: \begingroup \urlstyle{rm}\Url}\fi

\bibitem[Airoldi et~al.(2008)Airoldi, Blei, Fienberg, and
  Xing]{airoldi2008mixed}
E.~M. Airoldi, D.~Blei, S.~Fienberg, and E.~Xing.
\newblock Mixed membership stochastic blockmodels.
\newblock \emph{Advances in neural information processing systems}, 21, 2008.

\bibitem[Amini et~al.(2013)Amini, Chen, Bickel, and Levina]{amini2013pseudo}
A.~A. Amini, A.~Chen, P.~J. Bickel, and E.~Levina.
\newblock {Pseudo-likelihood methods for community detection in large sparse
  networks}.
\newblock \emph{The Annals of Statistics}, 41\penalty0 (4):\penalty0 2097 --
  2122, 2013.
\newblock \doi{10.1214/13-AOS1138}.
\newblock URL \url{https://doi.org/10.1214/13-AOS1138}.

\bibitem[Aoki et~al.(2012)Aoki, Hara, and Takemura]{AHT2012}
S.~Aoki, H.~Hara, and A.~Takemura.
\newblock \emph{Markov bases in algebraic statistics}, volume 199.
\newblock Springer Science \& Business Media, 2012.

\bibitem[Banerjee and Ma(2017)]{BanerjeeMa2017OptimalTests}
D.~Banerjee and Z.~Ma.
\newblock Optimal hypothesis testing for stochastic block models with growing
  degrees.
\newblock \emph{arXiv preprint arXiv:1705.05305}, 2017.

\bibitem[Barndorff-Nielsen(2014)]{barndorff2014information}
O.~Barndorff-Nielsen.
\newblock \emph{Information and exponential families: in statistical theory}.
\newblock John Wiley \& Sons, 2014.

\bibitem[Bickel et~al.(2013)Bickel, Choi, Chang, and Zhang]{BickelMLEsbm}
P.~Bickel, D.~Choi, X.~Chang, and H.~Zhang.
\newblock {Asymptotic normality of maximum likelihood and its variational
  approximation for stochastic blockmodels}.
\newblock \emph{The Annals of Statistics}, 41\penalty0 (4):\penalty0 1922 --
  1943, 2013.
\newblock \doi{10.1214/13-AOS1124}.
\newblock URL \url{https://doi.org/10.1214/13-AOS1124}.

\bibitem[Bickel and Chen(2009)]{bickel2009nonparametric}
P.~J. Bickel and A.~Chen.
\newblock A nonparametric view of network models and newman--girvan and other
  modularities.
\newblock \emph{Proceedings of the National Academy of Sciences}, 106\penalty0
  (50):\penalty0 21068--21073, 2009.

\bibitem[Brown(1986)]{brown1986}
L.~D. Brown.
\newblock \emph{Fundamentals of Statistical Exponential Families}, volume~9 of
  \emph{Monograph Series}.
\newblock IMS Lecture Notes, Hayward, CA, 1986.

\bibitem[Carnegie et~al.(2015)Carnegie, Krivitsky, Hunter, and
  Goodreau]{carnegie2015approximation}
N.~B. Carnegie, P.~N. Krivitsky, D.~R. Hunter, and S.~M. Goodreau.
\newblock An approximation method for improving dynamic network model fitting.
\newblock \emph{Journal of Computational and Graphical Statistics}, 24\penalty0
  (2):\penalty0 502--519, 2015.

\bibitem[Casanellas et~al.(2020)Casanellas, Petrovi{\'c}, and Uhler]{ARSIA}
M.~Casanellas, S.~Petrovi{\'c}, and C.~Uhler.
\newblock Algebraic statistics in practice: applications to networks.
\newblock \emph{Annual Review of Statistics and Its Application}, 7:\penalty0
  227--250, 2020.

\bibitem[Chatterjee et~al.(2011)Chatterjee, Diaconis, and Sly]{CDS11}
S.~Chatterjee, P.~Diaconis, and A.~Sly.
\newblock Random graphs with a given degree sequence.
\newblock \emph{The Annals of Applied Probability}, pages 1400--1435, 2011.

\bibitem[Diaconis and Sturmfels(1998)]{DS98}
P.~Diaconis and B.~Sturmfels.
\newblock Algebraic algorithms for sampling from conditional distributions.
\newblock \emph{The Annals of statistics}, 26\penalty0 (1):\penalty0 363--397,
  1998.

\bibitem[Erd{\H{o}}s et~al.(1960)Erd{\H{o}}s, R{\'e}nyi,
  et~al.]{erdos1961evolution}
P.~Erd{\H{o}}s, A.~R{\'e}nyi, et~al.
\newblock On the evolution of random graphs.
\newblock \emph{Publ. math. inst. hung. acad. sci}, 5\penalty0 (1):\penalty0
  17--60, 1960.

\bibitem[Fienberg(2012)]{Fienberg2012:briefHistoryNtwks}
S.~E. Fienberg.
\newblock A brief history of statistical models for network analysis and open
  challenges.
\newblock \emph{Journal of Computational and Graphical Statistics}, 21\penalty0
  (4):\penalty0 825--839, 2012.

\bibitem[Fienberg and Wasserman(1981)]{fienberg1981categorical}
S.~E. Fienberg and S.~S. Wasserman.
\newblock Categorical data analysis of single sociometric relations.
\newblock \emph{Sociological methodology}, 12:\penalty0 156--192, 1981.

\bibitem[Fienberg et~al.(1985)Fienberg, Meyer, and
  Wasserman]{fienberg1985statistical}
S.~E. Fienberg, M.~M. Meyer, and S.~S. Wasserman.
\newblock Statistical analysis of multiple sociometric relations.
\newblock \emph{Journal of the American Statistical Association}, 80\penalty0
  (389):\penalty0 51--67, 1985.

\bibitem[Fienberg et~al.(2007)Fienberg, Hersh, Rinaldo, and
  Zhou]{FHRZ07-MLEforLatent}
S.~E. Fienberg, P.~Hersh, A.~Rinaldo, and Y.~Zhou.
\newblock \emph{Maximum Likelihood Estimation in Latent Class Models For
  Contingency Table Data}, volume Algebraic and Geometric Methods in
  Statistics.
\newblock Cambridge University Press, 2007.

\bibitem[Fienberg et~al.(2011)Fienberg, Petrovi{\'c}, and
  Rinaldo]{SteveAleMe-holland}
S.~E. Fienberg, S.~Petrovi{\'c}, and A.~Rinaldo.
\newblock Algebraic statistics for p 1 random graph models: Markov bases and
  their uses.
\newblock In \emph{Looking Back: Proceedings of a Conference in Honor of Paul
  W. Holland}, pages 21--38. Springer, 2011.

\bibitem[Frank and Strauss(1986)]{frank1986markov}
O.~Frank and D.~Strauss.
\newblock Markov graphs.
\newblock \emph{Journal of the American Statistical Association}, 81\penalty0
  (395):\penalty0 832--842, 1986.

\bibitem[Fu et~al.(2009)Fu, Song, and Xing]{fu2009dynamic}
W.~Fu, L.~Song, and E.~P. Xing.
\newblock Dynamic mixed membership blockmodel for evolving networks.
\newblock In \emph{Proceedings of the 26th annual international conference on
  machine learning}, pages 329--336, 2009.

\bibitem[Gelfand and Ghosh(1998)]{gelfand1998model}
A.~E. Gelfand and S.~K. Ghosh.
\newblock Model choice: a minimum posterior predictive loss approach.
\newblock \emph{Biometrika}, 85\penalty0 (1):\penalty0 1--11, 1998.

\bibitem[Gelman et~al.(1996)Gelman, Meng, and Stern]{gelman1996posterior}
A.~Gelman, X.-L. Meng, and H.~Stern.
\newblock Posterior predictive assessment of model fitness via realized
  discrepancies.
\newblock \emph{Statistica Sinica}, pages 733--760, 1996.

\bibitem[Geng et~al.(2019)Geng, Bhattacharya, and Pati]{geng2016probabilistic}
J.~Geng, A.~Bhattacharya, and D.~Pati.
\newblock Probabilistic community detection with unknown number of communities.
\newblock \emph{Journal of the American Statistical Association}, 114\penalty0
  (526):\penalty0 893--905, 2019.

\bibitem[Geyer(2009)]{GeyerExpFamFan}
C.~J. Geyer.
\newblock {Likelihood inference in exponential families and directions of
  recession}.
\newblock \emph{Electronic Journal of Statistics}, 3\penalty0 (none):\penalty0
  259 -- 289, 2009.
\newblock \doi{10.1214/08-EJS349}.
\newblock URL \url{https://doi.org/10.1214/08-EJS349}.

\bibitem[Ghosh et~al.(2020)Ghosh, Pati, and Bhattacharya]{pati2015optimal}
P.~Ghosh, D.~Pati, and A.~Bhattacharya.
\newblock Posterior contraction rates for stochastic block models.
\newblock \emph{Sankhya A}, 82:\penalty0 448--476, 2020.

\bibitem[Gilbert(1959)]{Gilbert}
E.~N. Gilbert.
\newblock Random graphs.
\newblock \emph{The Annals of Mathematical Statistics}, 30\penalty0
  (4):\penalty0 1141--1144, 1959.

\bibitem[Girvan and Newman(2002)]{girvan2002community}
M.~Girvan and M.~E. Newman.
\newblock Community structure in social and biological networks.
\newblock \emph{Proceedings of the national academy of sciences}, 99\penalty0
  (12):\penalty0 7821--7826, 2002.

\bibitem[Glasser et~al.(2013)Glasser, Sotiropoulos, Wilson, Coalson, Fischl,
  Andersson, Xu, Jbabdi, Webster, Polimeni, et~al.]{Glasser2013105}
M.~F. Glasser, S.~N. Sotiropoulos, J.~A. Wilson, T.~S. Coalson, B.~Fischl,
  J.~L. Andersson, J.~Xu, S.~Jbabdi, M.~Webster, J.~R. Polimeni, et~al.
\newblock The minimal preprocessing pipelines for the human connectome project.
\newblock \emph{Neuroimage}, 80:\penalty0 105--124, 2013.

\bibitem[Goldenberg et~al.(2010)Goldenberg, Zheng, Fienberg, Airoldi,
  et~al.]{goldenberg2010survey}
A.~Goldenberg, A.~X. Zheng, S.~E. Fienberg, E.~M. Airoldi, et~al.
\newblock A survey of statistical network models.
\newblock \emph{Foundations and Trends{\textregistered} in Machine Learning},
  2\penalty0 (2):\penalty0 129--233, 2010.

\bibitem[Grayson and Stillman(2002)]{M2}
D.~R. Grayson and M.~E. Stillman.
\newblock Macaulay2, a software system for research in algebraic geometry,
  2002.
\newblock URL \url{http://www.math.uiuc.edu/Macaulay2/}.

\bibitem[Gross et~al.(2017)Gross, Petrovi{\'c}, and Stasi]{GPS16}
E.~Gross, S.~Petrovi{\'c}, and D.~Stasi.
\newblock Goodness of fit for log-linear network models: dynamic markov bases
  using hypergraphs.
\newblock \emph{Annals of the Institute of Statistical Mathematics},
  69\penalty0 (3):\penalty0 673--704, 2017.

\bibitem[Gross et~al.(2021)Gross, Petrovi{\'c}, and Stasi]{GPS21+}
E.~Gross, S.~Petrovi{\'c}, and D.~Stasi.
\newblock Goodness of fit for log-linear ergms.
\newblock \emph{arXiv preprint arXiv:2104.03167}, 2021.

\bibitem[Haberman(1988)]{Haberman88}
S.~J. Haberman.
\newblock A warning on the use of chi-squared statistics with frequency tables
  with small expected cell counts.
\newblock \emph{Journal of the American Statistical Association}, 83\penalty0
  (402):\penalty0 555--560, 1988.

\bibitem[Hoff et~al.(2002)Hoff, Raftery, and Handcock]{hoff2002latent}
P.~D. Hoff, A.~E. Raftery, and M.~S. Handcock.
\newblock Latent space approaches to social network analysis.
\newblock \emph{Journal of the American Statistical Association}, 97\penalty0
  (460):\penalty0 1090--1098, 2002.

\bibitem[Holland and Leinhardt(1981)]{holland1981exponential}
P.~W. Holland and S.~Leinhardt.
\newblock An exponential family of probability distributions for directed
  graphs.
\newblock \emph{Journal of the American Statistical Association}, 76\penalty0
  (373):\penalty0 33--50, 1981.

\bibitem[Holland et~al.(1983)Holland, Laskey, and
  Leinhardt]{holland1983stochastic}
P.~W. Holland, K.~B. Laskey, and S.~Leinhardt.
\newblock Stochastic blockmodels: First steps.
\newblock \emph{Social networks}, 5\penalty0 (2):\penalty0 109--137, 1983.

\bibitem[Hunter et~al.(2008{\natexlab{a}})Hunter, Goodreau, and
  Handcock]{HunterGoF}
D.~R. Hunter, S.~M. Goodreau, and M.~S. Handcock.
\newblock Goodness of fit of social network models.
\newblock \emph{Journal of the American Statistical Association}, 103\penalty0
  (481):\penalty0 248--258, 2008{\natexlab{a}}.

\bibitem[Hunter et~al.(2008{\natexlab{b}})Hunter, Handcock, Butts, Goodreau,
  and Morris]{hunter2008ergm}
D.~R. Hunter, M.~S. Handcock, C.~T. Butts, S.~M. Goodreau, and M.~Morris.
\newblock ergm: A package to fit, simulate and diagnose exponential-family
  models for networks.
\newblock \emph{Journal of statistical software}, 24\penalty0 (3):\penalty0
  nihpa54860, 2008{\natexlab{b}}.

\bibitem[Karrer and Newman(2011)]{karrer2011stochastic}
B.~Karrer and M.~E. Newman.
\newblock Stochastic blockmodels and community structure in networks.
\newblock \emph{Physical review E}, 83\penalty0 (1):\penalty0 016107, 2011.

\bibitem[Karwa and Petrovi\'c(2016)]{KP:AOAS}
V.~Karwa and S.~Petrovi\'c.
\newblock Coauthorship and citation networks for statisticians: Comment.
  invited comment on the paper by {J}in and {J}i.
\newblock \emph{Annals of Applied Statistics}, 10\penalty0 (4):\penalty0
  1827--1834, 2016.

\bibitem[Karwa and Slavkovi{\'c}(2012)]{SesaVishesBetaPrivacy}
V.~Karwa and A.~B. Slavkovi{\'c}.
\newblock Differentially private graphical degree sequences and synthetic
  graphs.
\newblock In \emph{Privacy in Statistical Databases: UNESCO Chair in Data
  Privacy, International Conference, PSD 2012, Palermo, Italy, September 26-28,
  2012. Proceedings}, pages 273--285. Springer, 2012.

\bibitem[Karwa and Slavković(2016)]{SesaVishesBetaPrivacyAOS}
V.~Karwa and A.~Slavković.
\newblock {Inference using noisy degrees: Differentially private $\beta$-model
  and synthetic graphs}.
\newblock \emph{The Annals of Statistics}, 44\penalty0 (1):\penalty0 87 -- 112,
  2016.
\newblock \doi{10.1214/15-AOS1358}.
\newblock URL \url{https://doi.org/10.1214/15-AOS1358}.

\bibitem[Karwa et~al.(2023)Karwa, Pati, Petrovi\'c, Solus, Alexeev, Rai\v{c},
  Wilburne, Williams, and Yan]{mrc2016-supplement}
V.~Karwa, D.~Pati, S.~Petrovi\'c, L.~Solus, N.~Alexeev, M.~Rai\v{c},
  D.~Wilburne, R.~Williams, and B.~Yan.
\newblock Supplementary material to ``{M}onte {C}arlo goodness-of-fit tests for
  degree corrected and related stochastic blockmodels''.
\newblock \emph{Supplementary material}, 2023.

\bibitem[Kolaczyk(2017)]{KolaczykBook2017}
E.~D. Kolaczyk.
\newblock \emph{Topics at the Frontier of Statistics and Network Analysis:(re)
  visiting the Foundations}.
\newblock Cambridge University Press, 2017.

\bibitem[Kolaczyk and Krivitsky(2015)]{Kolaczyk-EffectiveSampleSize}
E.~D. Kolaczyk and P.~N. Krivitsky.
\newblock On the question of effective sample size in network modeling: An
  asymptotic inquiry.
\newblock \emph{Statistical science: a review journal of the Institute of
  Mathematical Statistics}, 30\penalty0 (2):\penalty0 184, 2015.

\bibitem[Lauritzen(1996)]{lauritzen1996graphical}
S.~L. Lauritzen.
\newblock \emph{Graphical models}, volume~17.
\newblock Clarendon Press, 1996.

\bibitem[Lei(2016)]{lei2016goodness}
J.~Lei.
\newblock {A goodness-of-fit test for stochastic block models}.
\newblock \emph{The Annals of Statistics}, 44\penalty0 (1):\penalty0 401 --
  424, 2016.
\newblock \doi{10.1214/15-AOS1370}.
\newblock URL \url{https://doi.org/10.1214/15-AOS1370}.

\bibitem[Mahadev and Peled(1995)]{MP95}
N.~V. Mahadev and U.~N. Peled.
\newblock \emph{Threshold graphs and related topics}.
\newblock Elsevier, 1995.

\bibitem[Matias and Miele(2017)]{matias2015statistical}
C.~Matias and V.~Miele.
\newblock Statistical clustering of temporal networks through a dynamic
  stochastic block model.
\newblock \emph{Journal of the Royal Statistical Society Series B: Statistical
  Methodology}, 79\penalty0 (4):\penalty0 1119--1141, 2017.

\bibitem[Meng(1994)]{meng1994posterior}
X.-L. Meng.
\newblock Posterior predictive $ p $-values.
\newblock \emph{The Annals of Statistics}, 22\penalty0 (3):\penalty0
  1142--1160, 1994.

\bibitem[Newman and Reinert(2016)]{NewmanReinert2016EstimatingNumBlocks}
M.~E. Newman and G.~Reinert.
\newblock Estimating the number of communities in a network.
\newblock \emph{Physical review letters}, 117\penalty0 (7):\penalty0 078301,
  2016.

\bibitem[Nowicki and Snijders(2001)]{nowicki2001estimation}
K.~Nowicki and T.~A.~B. Snijders.
\newblock Estimation and prediction for stochastic blockstructures.
\newblock \emph{Journal of the American statistical association}, 96\penalty0
  (455):\penalty0 1077--1087, 2001.

\bibitem[Peng and Carvalho(2016)]{peng2013bayesian}
L.~Peng and L.~Carvalho.
\newblock {Bayesian degree-corrected stochastic blockmodels for community
  detection}.
\newblock \emph{Electronic Journal of Statistics}, 10\penalty0 (2):\penalty0
  2746 -- 2779, 2016.
\newblock \doi{10.1214/16-EJS1163}.
\newblock URL \url{https://doi.org/10.1214/16-EJS1163}.

\bibitem[Petrovi\'c(2019)]{WhatIsMB}
S.~Petrovi\'c.
\newblock What is... {A} {M}arkov basis?
\newblock \emph{Notices of the American Mathematical Society}, 66\penalty0
  (7):\penalty0 1088---1092, 2019.

\bibitem[Petrovic et~al.(2010)Petrovic, Rinaldo, and Fienberg]{PRF10}
S.~Petrovic, A.~Rinaldo, and S.~E. Fienberg.
\newblock Algebraic statistics for a directed random graph model with
  reciprocation.
\newblock \emph{Algebraic methods in statistics and probability II},
  516:\penalty0 261--283, 2010.

\bibitem[Qin and Rohe(2013)]{qin2013regularized}
T.~Qin and K.~Rohe.
\newblock Regularized spectral clustering under the degree-corrected stochastic
  blockmodel.
\newblock \emph{Advances in neural information processing systems}, 26, 2013.

\bibitem[Rai\v{c}(2019)]{MatejaPhD}
M.~Rai\v{c}.
\newblock \emph{Atomic test for goodness of fit of network models}.
\newblock PhD thesis, University of Illinois at Chicago, 2019.

\bibitem[Rinaldo et~al.(2010)Rinaldo, Petrovi{\'c}, and Fienberg]{RPF:10}
A.~Rinaldo, S.~Petrovi{\'c}, and S.~E. Fienberg.
\newblock On the existence of the mle for a directed random graph network model
  with reciprocation.
\newblock \emph{arXiv preprint arXiv:1010.0745}, 2010.

\bibitem[Rinaldo et~al.(2013)Rinaldo, Petrović, and Fienberg]{RPF:11}
A.~Rinaldo, S.~Petrović, and S.~E. Fienberg.
\newblock {Maximum lilkelihood estimation in the $\beta$-model}.
\newblock \emph{The Annals of Statistics}, 41\penalty0 (3):\penalty0 1085 --
  1110, 2013.
\newblock \doi{10.1214/12-AOS1078}.
\newblock URL \url{https://doi.org/10.1214/12-AOS1078}.

\bibitem[Robins et~al.(2007)Robins, Pattison, Kalish, and Lusher]{Robins2007}
G.~Robins, P.~Pattison, Y.~Kalish, and D.~Lusher.
\newblock An introduction to exponential random graph (p*) models for social
  networks.
\newblock \emph{Social networks}, 29\penalty0 (2):\penalty0 173--191, 2007.

\bibitem[Yan and Sarkar(2021)]{yan2016convex}
B.~Yan and P.~Sarkar.
\newblock Covariate regularized community detection in sparse graphs.
\newblock \emph{Journal of the American Statistical Association}, 116\penalty0
  (534):\penalty0 734--745, 2021.
\newblock \doi{10.1080/01621459.2019.1706541}.
\newblock URL \url{https://doi.org/10.1080/01621459.2019.1706541}.

\bibitem[Yang(2015)]{Xiaolin2015thesis}
X.~Yang.
\newblock \emph{Social network modeling and the evaluation of structural
  similarity for community detection}.
\newblock PhD thesis, Ph. D. thesis, Carnegie Mellon University, 2015.

\bibitem[Zachary(1977)]{zachary1977information}
W.~W. Zachary.
\newblock An information flow model for conflict and fission in small groups.
\newblock \emph{Journal of anthropological research}, 33\penalty0 (4):\penalty0
  452--473, 1977.

\bibitem[Zhang et~al.(2019)Zhang, Allen, Zhu, and Dunson]{zhang2019tensor}
Z.~Zhang, G.~I. Allen, H.~Zhu, and D.~Dunson.
\newblock Tensor network factorizations: Relationships between brain structural
  connectomes and traits.
\newblock \emph{Neuroimage}, 197:\penalty0 330--343, 2019.

\end{thebibliography}

\listoffigures

\end{document}